\documentclass[12pt,a4paper]{article}
\pdfoutput=1
\usepackage{jheppub}
\usepackage{amsmath}
\usepackage{amsfonts}
\usepackage{mathtools}
\usepackage{amssymb}
\usepackage{caption}
\usepackage{subcaption}
\usepackage[utf8]{inputenc}
\allowdisplaybreaks[2]
\def\be{\begin{equation}}
\def\ee{\end{equation}}
\def\bea{\begin{eqnarray}}
\def\eea{\end{eqnarray}}
\title{Local quenches and quantum chaos  from higher spin perturbations}
\author{Justin R. David ${}^{a}$, Surbhi Khetrapal ${}^{a}$, S. Prem Kumar${}^{b}$ }
\affiliation{${}^{a}$ Centre for High Energy Physics, Indian Institute of Science,\\
C. V. Raman Avenue, Bangalore 560012, India. \\
${}^{b}$ Department of Physics, Swansea University, \\
Singleton Park, Swansea SA2 8PP, UK. }
\emailAdd{justin, surbhi@chep.iisc.ernet.in, s.p.kumar@swansea.ac.uk}
\abstract{ We study local quenches in 1+1 dimensional conformal field theories at large-$c$ by 
operators carrying higher spin charge. Viewing such states as solutions in Chern-Simons theory, representing 
infalling massive particles  with spin-three charge in the BTZ background, we use the Wilson line 
prescription to compute the single-interval entanglement entropy (EE) and scrambling time 
following the quench. We find that the change in  EE is finite (and real) only if the spin-three 
charge $q$ is bounded by the energy of the perturbation $E$, as $|q|/c < E^2/c^2$. 
We show that the Wilson line/EE correlator deep in the quenched regime 
and its expansion for small quench widths overlaps with the Regge limit for chaos of the 
out-of-time-ordered correlator.
 We further find that the scrambling time for the two-sided mutual information between 
 two intervals in the thermofield  double state 
increases  with increasing spin-three charge, diverging when the bound is saturated. For larger values of 
the charge, the scrambling time is shorter than for pure gravity and controlled by the 
spin-three Lyapunov exponent $4\pi/\beta$. In a CFT with  higher spin chemical potential, dual to a 
higher spin black hole, we find that the chemical potential must be bounded to 
 ensure  that the mutual information is  a concave  function of time and entanglement speed 
is less than the speed of light. In this  case, a 
quench with zero higher spin charge yields the same Lyapunov exponent as pure Einstein gravity.
}
\begin{document}
\maketitle
\flushbottom
\section{Introduction}
The incorporation of ideas from quantum chaos is an exciting development in the study of real time dynamics of quantum field theories (QFTs) and its implications for gravitational systems which are holographically dual to them \cite{Shenker:2013pqa, Maldacena:2015waa, Shenker:2013yza,Shenker:2014cwa, Roberts:2014isa, Roberts:2014ifa, Kitaev}.  In particular, the bound proposed in \cite{Shenker:2013pqa,Maldacena:2015waa} identifies black holes in Einstein gravity as possessing the fastest possible scrambling time which controls the  onset of chaotic exponential decay of correlators in large-$N$ quantum field theories  with gravity duals. This proposal, which is fascinating in its own right, also has implications for theories of gravity  which are potentially dual to conformal field theories (CFTs) at large-$N$ or large central charge $c$.

This work was motivated in part by the observations of  \cite{Perlmutter:2016pkf} wherein restrictions on theories of gravity containing higher spin fields were deduced by analysing the temporal behaviour of 
out-of-time-ordered (OTO) correlators. In particular, it was argued that for CFTs with only a finite number of higher spin currents, OTO correlators can exhibit unbounded growth in time and violation of the proposed lower bound \cite{Shenker:2013pqa,Maldacena:2015waa} on scrambling times. The conclusions were drawn by computing correlators of two heavy (H) and two light (L) operators using the semiclassical $W_N$ conformal blocks at large $c$ \cite{Hartman:2013mia, Asplund:2014coa, deBoer:2014sna, Besken:2016ooo}. 

In this paper we focus attention on the temporal behaviour of entanglement entropies in ${\rm CFT}_2$ following a local quench \cite{Calabrese:2007mtj, Nozaki:2013wia, Nozaki:2014hna, He:2014mwa, Caputa:2014vaa, paper1} by  a CFT primary ${\cal O}$ carrying higher spin charge. In the local quench the equilibrium density matrix $\rho_\beta\,=\,e^{-\beta H}$ of the CFT at some  temperature $\beta^{-1}$ is perturbed locally at time $t=0$ :
\be
\rho_\beta \,\to\, {\cal O}(i\epsilon)\,\rho_\beta\,{\cal O}^\dagger(-i\epsilon)\,,
\ee 
and the state then evolved in time. The parameter $\epsilon$ controls the width of the excitation produced by the perturbation.
Our approach is to adapt the holographic ${\rm AdS}_3$ calculation in Einstein gravity of \cite{Caputa1, Caputa2} to  higher spin theory; specifically, the ${\rm SL}(3,{\mathbb R})\times {\rm SL}(3,{\mathbb R})$ Chern-Simons theory which extends Einstein gravity to include a spin three field. The local quench of the CFT in a thermal state is described in  Einstein gravity by 
the backreacted geometry due to a particle (conical deficit) freely falling into the BTZ black hole. This geometry can be obtained by a coordinate transformation and boost \cite{Nozaki:2013wia, Caputa1} on a static conical deficit state in global ${\rm AdS}_3$.  The bulk diffeomorphism acts as a conformal transformation on the boundary. In the higher spin theory, formulated as Chern-Simons theory, there is no gauge-invariant notion of geometry. Boundary CFT entanglement entropies are computed by Chern-Simons Wilson lines \cite{deBjottar,amcasiq} anchored to the endpoints of intervals on the conformal boundary of ${\rm AdS}_3$. We compute such Wilson lines in a static, spin-three charged, conical deficit state, characterised by a flat Chern-Simons connection, and act on the result by the same boundary conformal transformation which maps the uncharged deficits to infalling massive particles in the BTZ background. This is then interpreted as a finite width local quench by an operator carrying spin-three charge, with the CFT originally in  the zero charge thermal ensemble\footnote{ It is expected that the Wilson line/EE computation should be equivalent to the evaluation of  HHLL correlators using conformal blocks at large $c$ \cite{deBoer:2014sna, Besken:2016ooo}.}. The generalisation of 
this approach to local quenches in a grand canonical ensemble for higher spin charge is not straightforward.  We can, however, use existing CFT results for finite spin-three chemical potential \cite{universalcft} and properties of HHLL correlators to infer what happens in the grand canonical ensemble when the perturbing operator ${\cal O}$ carries no higher spin charge. Our main findings are summarised below:
\begin{itemize}
\item{The quench generated by the operator ${\cal O}$ with conformal dimension $\Delta_{\cal O}$ and spin-three quantum number\footnote{${\cal W}$ may be viewed as a dimensionless number appearing in the OPE of the spin three current $W$ with ${\cal O}$, assuming that the latter transforms as a primary under the spin three current
\be
W(x)\,{\cal O}(y,\bar y)\,\sim \,\frac{{\cal W}}{(x-y)^3}{\cal O}(y,\bar y)\,.
\ee
} ${\cal W}$ is a pulse of width $\epsilon\ll \beta$. It carries an energy density $\langle T_{00}\rangle \sim \Delta_{\cal O}/\epsilon^2$ and spin three charge density $\langle W\rangle \sim {\cal W}/\epsilon^3$. In order to keep the total energy and charge of the pulse fixed in the small $\epsilon$ limit, we take $\Delta_{\cal O}\,=\, \epsilon \frac{E}{\pi}$ and 
${\cal W}\,=\,\epsilon^2 \frac{4q}{\pi^2}$. Furthermore, to ensure that the effect of the quench remains non-vanishing, we also need to keep $E/c$ and $q/c$ fixed in the large-$c$ limit. In this double-scaled limit, the  Wilson line computation of the single interval entanglement entropy following the quench then yields a finite ``jump" in the entanglement entropy $\Delta S_{\rm EE}$ when the pulse enters the interval of interest.  We find that with non-zero spin three charge, $\Delta S_{\rm EE}$ is not positive definite, and furthermore, remains real and finite only if the condition,
\be
\frac{|q|}{c}\, < \,\frac{E^2}{c^2}
\ee
is satisfied.
 }
 \item{The temporal regime wherein the excitation is deep in the interval of interest, and $\Delta S_{\rm EE}$ has saturated, can be accessed by a small width expansion of the Wilson line correlator in the double-scaled limit explained above. We find that this expansion when expressed in terms of the conformal cross-ratio $z$, coincides with the expansion of the OTO correlator in the Regge limit for chaos in \cite{Perlmutter:2016pkf}. Although it is no  surprise that the Wilson line coincides with the HHLL correlator in the large $c$ semiclassical limit, it is interesting that the two physically distinct phenomena originate from the same expansion of the correlator when expressed in terms of the appropriate conformal cross-ratio $z$. Put differently, while $\Delta S_{\rm EE}$ and the late time OTO correlator have different time dependence, both are determined by the small $z$ expansion of a particular branch of the same analytic function of $z$. }
 \item{We use the Wilson line correlator to calculate the scrambling time following the approach of \cite{Caputa2}. Specifically, this involves taking the CFT in the thermofield double state and calculating the mutual information of two intervals, one on each copy of the CFT, in the presence of the local quench perturbation introduced on one copy. Again, we use the ${\rm SL}(3,{\mathbb R})$ Wilson line for the charged conical deficit to calculate the mutual information in the thermofield double state. This is achieved by identifying the correct conformal transformations which map boundary points in global ${\rm AdS}_3$ to the two sets of boundary points in the Kruskal extension of the eternal BTZ black hole \cite{Caputa2}. The mutual information then receives connected and disconnected bulk contributions, and the time at which it vanishes is identified as the scrambling time \eqref{t*} which is evaluated in the double-scaled limit we have described previously. The result has the feature that for small spin three charge $q$, the scrambling time {\em increases} beyond its pure Einstein gravity value until it diverges precisely when ${|q|}/{c}\, = \,{E^2}/{c^2}$. Interestingly, the formula \eqref{t*} continues to makes sense also above this bound, so that in the limit that the spin three charge dominates, the scrambling time is shorter than pure gravity and the associated Lyapunov exponent is $4\pi/\beta$, in line with the arguments of \cite{Perlmutter:2016pkf} where a Lyapunov exponent $2\pi (N-1)/\beta$ was obtained in the presence of a spin-$N$ charge.}
 
\item {A question of particular interest is how the computations above 
(and the arguments of \cite{Perlmutter:2016pkf}) generalise to the situation where the 
CFT is held at a chemical potential for higher spin charge. One reason this is nontrivial from a 
bulk perspective is that, in the absence of an invariant geometrical picture in the 
Chern-Simons formulation, it is not known whether the putative backreacted shockwave 
solution in the higher spin black hole background \cite{Gutperle:2011kf} can be 
obtained by systematically  transforming a conical deficit solution. We do not attempt this 
generalisation in this work. Instead, we focus our attention on two questions which can both 
be answered with currently known CFT and bulk results at finite spin three chemical potential. 
The first of these is to consider the thermofield double state for the 
spin three black hole ({\em without} any external perturbation or quench) of \cite{Gutperle:2011kf} 
and find the time evolution of mutual information for two intervals in the 
two different copies under forward time evolution of both 
copies \cite{Calabrese:2005in, Hartman:2013qma}. This can be obtained without 
explicit knowledge of the Kruskal extension of the spin 
three black hole \cite{Castro:2016ehj}, by simply analytically continuing CFT results of \cite{universalcft} 
and by doing the same to the time coordinates of the endpoints of bulk Wilson lines to go 
between the two copies of the thermofield double. Using the {\em holomorphic} Wilson 
line \cite{deBjottar} which agrees with CFT results \cite{universalcft}, we find that 
there is a critical value of the spin three chemical potential
beyond which the mutual information ceases to be a concave 
function of time and simultaneously, the speed of growth of entanglement entropy of the intervals exceeds unity. 
%This suggests a potentially pathological feature of the black hole 
%solution with finite spin three chemical potential. 
Moving to the situation with a local quench, 
using the known results for CFT entanglement entropy at finite $\mu$ and properties of 
HHLL correlators where the heavy operator $\cal O$ carries no spin three charge, we 
argue that the Lyapunov exponent retains its value in Einstein gravity while the scrambling 
time receives some $\mu$-dependent corrections.}
\end{itemize}

The paper is organised as follows.
In section \ref{sec:one}, we review the Wilson line prescription for evaluating entanglement entropy in ${\rm SL}(2,{\mathbb R})$ Chern Simons theory in the presence of a local quench.  We point out the connection of the quenched regime in the small width expansion, with the OTO correlator and its Regge limit for chaos.
In section \ref{sec:sl3ee} we repeat the exercise of for the quench with spin three charge. We further find the scrambling time by computing the two-sided mutual information.  We
study the situation in section \ref{HS_CD} when the local quench is generated by an ensemble of operators carrying higher spin charge.  Section \ref{scrambling_CFT} is devoted to aspects of the CFT dynamics in the presence of higher spin chemical potential.
In a fairly extensive appendix, we present detailed clarifications, derivations of various technical points in the text. We also include two sections which outline aspects of the bulk quench and scrambling calculations in Chern-Simons language.

\section {Wilson lines, local quenches and OTO correlators}
\label{sec:one}
The primary objects of our interest are  correlators 
involving heavy (H) and  light (L) operators which yield the time evolution of entanglement in CFT$_2$ at finite temperature in the presence of a local excitation. In CFT$_2$, entanglement/R\'enyi  entropies are computed by the insertion of local twist fields. The latter are ``light" in the limit that the number of replicas approaches unity. 

In the limit of large central charge $c$, when a dual gravity description becomes appropriate, the entanglement entropy is computed by the Ryu-Takayanagi prescription \cite{Ryu:2006bv}. In the case of ${\rm AdS}_3/{\rm CFT}_2$ duality, gravity and its higher spin generalisations are naturally recast in the language of ${\rm SL}(N, {\mathbb R})\times {\rm SL}(N,{\mathbb R})$ Chern-Simons theory. The Ryu-Takayanagi prescription then generalises to a Wilson line in an appropriate representation anchored at the endpoints of the interval whose entanglement entropy is being evaluated \cite{deBjottar, amcasiq}.

The four-point correlators of interest are of the form
\begin{equation}
F(x_1,x_2,x_3,x_4)\,=\,\langle {\cal O}^\dagger(x_1,\bar x_1)\,{\cal T}(x_2, \bar x_2)\,
\widetilde {\cal T}(x_3, \bar x_3)\,{\cal O}(x_4, \bar x_4)\rangle\,,
\label{4pt}
\end{equation}
where ${\cal O}$ is the heavy operator and ${\cal T}$ represents the light operator.
Such correlators are natural when one considers the time evolution of quantum entanglement after a ``quench" by some (heavy) local operator ${\cal O}(x,t)$ \cite{He:2014mwa, Caputa1, Caputa2, paper1}. Following the conventions of \cite{Caputa2} for example, we may take
\bea
\label{positions}
&& x_1\,=\,-i\epsilon\,,\qquad\qquad\,\quad \bar x_1\,=\,+i\epsilon\,,\label{hhllpoints}\\\nonumber
&& x_2\,=\,\ell_1\, -\, t\,,\qquad\qquad \bar x_2\,=\,\ell_1\, +\,t\,,\\\nonumber
&& x_3\,=\,\ell_2\, -\, t\,,\qquad\qquad \bar x_3\,=\,\ell_2\, +\,t\,,\\\nonumber
&& x_4\,=\,i\epsilon\,,\qquad\qquad\qquad \bar x_4\,=\,-i\epsilon\,.
\eea
Here $(\ell_1, \ell_2)$ represent the spatial coordinates of the entangling interval of length $\ell\,\equiv\,\ell_2-\ell_1$, and $\epsilon >0 $ denotes the ``width" of the local quench \cite{Caputa1, paper1}.  In addition to controlling the physical width of the pulse set up by the local perturbation,  the width $\epsilon$ serves to regulate the operator product and, importantly, allows us to track changes in the temporal behaviour of the correlator when the excitation crosses the lightcone of the nearest endpoint of the interval.

In the bulk gravity dual picture (for  a large-$c$ CFT), the effect of the local quench is reproduced by a shockwave background generated by a massive particle freely falling from the ${\rm AdS}_3$ boundary towards the interior. The excitation about the thermal state in the CFT is represented by the particle falling towards the horizon of a BTZ black hole in the bulk. The width of the excitation $\epsilon$ is related to an appropriately defined coordinate distance of the point of release of the particle from the boundary of ${\rm AdS_3}$ \cite{Shenker:2013pqa, Caputa1}.

The four-point correlator of the type \eqref{4pt} which yields the single interval entanglement entropy in the locally quenched quantum state is  computed by a Wilson line in the asymptotically ${\rm AdS}_3$ shockwave background . Given the pair of flat connections $(A, \bar A)$ valued in ${\rm sl}(N,{\mathbb R})\oplus {\rm sl}(N,{\mathbb R})$, the entanglement entropy of a single interval (with endpoints $P$ and $Q$ on the conformal boundary of ${\rm AdS}_3$) is given by the Wilson line representation ${\cal R}$ joining the endpoints:
\bea
&&S_{\rm EE} (P,Q)\,=\,k_{\rm cs}\,\ln\left[\lim_{\rho_{P,Q}\to\infty}\,W_{\cal R}(P,Q)\right]\,,\label{wlee}\\\nonumber\\\nonumber
&& W_{\cal R}(P, Q)\,\equiv\,{\rm Tr}_{\cal R}\left[{\cal P}\exp\left(\int_P^Q\bar A\right)\,\exp\left(\int_Q^P A\right)\right]\,.
\eea
Here $k_{\rm cs}$ is the level of the Chern-Simons theory which is related to the central charge of the asymptotic ${\cal W}_N$ algebra:  
\be
c\,=\, N(N^2-1)\,k_{\rm cs}\,,
\ee
and $\rho_{P,Q}$ are the radial coordinates of the endpoints of the interval on the boundary, to be taken to infinity at the end, so that only the leading term in this limit is identified with the entanglement entropy. The representation ${\cal R}$ is fixed by requiring the high temperature limit of the entanglement entropy to agree with the thermal entropy of the interval \cite{deBjottar}.

 We are  assuming that gravity is principally embedded in the ${\rm sl}(N,{\mathbb R})\oplus {\rm sl}(N,{\mathbb R})$ algebra. Denoting the generators of the irreducible $N$-dimensional representation of ${\rm sl}(2,{\mathbb R})$ as $\{L_0, L_{\pm 1}\}$ with $[L_0,L_{\pm 1}]\,=\,\pm L_{\pm 1}$ and $[L_1, L_{-1}]\,=\, 2L_0$, the flat Chern-Simons connections may be represented in radial gauge as 
 \bea
&& A \,=\, b^{-1} db \,+\, b^{-1}\, a(x^+,\,x^-)\,b \,,\qquad
\bar{A} \, =\, b db^{-1} \,+\, b\, \bar{a}(x^+,\,x^-)\, b^{-1}\,,\label{radialgauge}
 \\\nonumber\\\nonumber
&&b(\rho)\,=\,e^{\rho L_0}\,. 
\eea
Here $x^\pm$ are lightcone coordinates on the boundary, which we will specify precisely below.  Given the connections $(A, \bar A)$, the spacetime metric is determined as,
\be
ds^2\,=\,\frac{1}{4\epsilon_N}{\rm Tr}\left(A\,-\,\bar A\right)^2\,,\qquad\qquad \epsilon_N\,=\,{\rm Tr} L_0^2
\,=\,\frac{1}{12}N(N^2-1)\,.
\ee

\subsection{${\rm SL}(2,\mathbb R)\times {\rm SL}(2,\mathbb R)$ Wilson line and local quench} \label{holomorphic_pres_review}
As a warmup exercise we first rederive the evolution of EE following a local quench in pure gravity \cite{Caputa1}, but using the Wilson line prescription for calculating holographic EE in a shockwave geometry.

\paragraph{Conical deficit:} In \cite{Caputa1}, the shockwave geometry in the  BTZ black hole background was obtained by considering a conical deficit state in global ${\rm AdS}_3$ and performing a coordinate transformation followed by a boost. The metric for the static conical deficit in ${\rm AdS_3}$ is given by:
\be
ds^2 = -(r^2+R^2-\delta)\,d\tau^2\, +\, \frac{R^2\, dr^2}{r^2+R^2-\delta}\,+\,r^2\, d\phi^2\,,\label{staticdeficit}
\ee
where $R$ is the AdS radius and $\phi \in [0,2\pi]$. The mass, $m$, of the particle producing the conical deficit is fixed in terms of  $\delta$  as,
\be
\delta\,=\, 8\,(G_N R) mR\,=\,\frac{24 }{c}\Delta_{\cal O}\,R^2\,.
\ee
The conical deficit geometry represents  a CFT state corresponding to an operator of  conformal dimension $\Delta_{\cal O}\,=\,(mR)$. In order to make contact with the Chern-Simons formulation, we rewrite the conical deficit metric  in terms of lightcone coordinates $\xi^\pm$ on the boundary, and a new radial coordinate $\rho$:
\begin{align}\label{tauphi_xpm}
r \,=\, R e^{-\rho}\left(e^{2\rho} - \frac{R^2 -\delta}{4R^2}\right)\,, \qquad \tau\, =\, \frac{1}{2} (\xi^+ + \xi^-)\,, \qquad \phi\, =\, \frac{1}{2} (\xi^+ - \xi^-).
\end{align}
%This yields the metric:
%\begin{align} \label{conical_metric}
%ds^2 = R^2\left[ d\rho^2 + \tfrac{\delta - R^2}{4R^2} \left((d\xi^+)^2+ (d\xi^-)^2\right)- \left(e^{2\rho}+\left(\tfrac{\delta - R^2}{4R^2}\right)^2 e^{-2\rho}\right)d\xi^+ d\xi^-\right].
%\end{align}
This yields  a special case of the general form of asymptotically ${\rm AdS}_3$ solutions \cite{Banados:1998gg},
\bea
&&R^{-2}\,ds^2\, = \,\left[d\rho^2\, +\right.
\label{sl2r_metric}\\\nonumber\\\nonumber
&& \left. +\,\frac{2\pi}{k_{\rm cs}}{\cal L}(\xi^+)\, (d\xi^+)^2\,+\, \frac{2\pi}{k_{\rm cs}}\bar {\cal L}(\xi^-)\, \left(d\xi^-\right)^2\,-\, \left(e^{2\rho}\,+\,\frac{4\pi^2}{k^2_{\rm cs}}{\cal L}(\xi^+) \bar{\cal L}(\xi^-) e^{-2\rho}\right)\,d\xi^+ d\xi^-\right]\,,
\eea
which follows from  the flat ${\rm sl}(2,{\mathbb R})$ Chern-Simons connections as defined in the radial gauge \eqref{radialgauge},
\be
a \,=\,\left(L_1 \,-\, \frac{2\pi {\cal L}(\xi^+)}{k_{\rm cs}}L_{-1} \right)d\xi^+\,,\qquad \bar a \,=\,-\left(L_{-1} \,-\, \frac{2\pi {\cal L}(\xi^-)}{k_{\rm cs}}L_1 \right)d\xi^-\,.
\ee
The conical deficit state is obtained by setting $2\pi {\cal L}\,=\, 2\pi\bar {\cal L}\,=\,k_{\rm cs}(\delta-R^2)/4R^2$.

 In \cite{Caputa1}, this conical deficit state was mapped to an exact solution describing a massive infalling particle in the BTZ geometry.
Physical observables in the quenched state can be obtained by application of the same map. Thus we consider the entanglement entropy of a single interval with endpoints $\xi_P^\pm$ and $\xi_Q^\pm$ in the boundary CFT. The radial, holographic coordinate of the two points are $\rho_P$ and $\rho_Q$, which will eventually be taken to infinity.

 Note that for the ${\rm SL}(2, {\mathbb R})$ Chern-Simons theory, the Wilson line which computes holographic entanglement entropy is in the defining or fundamental representation \cite{deBjottar}. In terms of  the matrices,
\be\label{g_def}
g \, =\, \exp\left(a\, \xi^+\right)\,b(\rho)\,,\qquad
\bar g \, =\, \exp\left(\bar a \,\xi^-\right)\, b^{-1}(\rho)\,,
\ee
the fundamental Wilson line connecting the two boundary points $P$ and $Q$ in the conical deficit state is,
\begin{align}\label{WL_g_def}
W_{\rm fund}(P,Q) = \mathrm{Tr}_{\rm fund}\left[\bar g^{-1}(P)\, \bar g(Q)\, g^{-1}(Q)\, g(P)\right].
\end{align}
The traces are easily evaluated and we find,
\bea \label{sl2wl}
&& W_{\rm fund}(P,Q)\,=\,2 \cosh (\rho_P\,-\,\rho_Q) \,\cosh\left(\sqrt{\tfrac{2\pi {\cal L}}{k_{\rm cs}}}\,\Delta \xi^+ \right)\,\cosh\left(\sqrt{\tfrac{2\pi\bar {\cal L}}{k_{\rm cs}}}\,\Delta \xi^+ \right)\\
&&-\,\left(\tfrac{1}{2\pi}\tfrac{k_{\rm cs}}{\sqrt{{\cal L} \bar{\cal L}}} \,e^{\rho_P+\rho_Q}\, +\, \tfrac{2\pi\sqrt{{\cal L} \bar {\cal L}}}{k_{\rm cs}}\, e^{-(\rho_P+\rho_Q)} \right)\,\sinh\left(\sqrt{\tfrac{{2\pi\cal L}}{k_{\rm cs}}}\,\Delta \xi^+ \right)\,\sinh\left(\sqrt{\tfrac{2\pi\bar {\cal L}}{k_{\rm cs}}}\,\Delta \xi^- \right),\nonumber
\eea
where $\Delta \xi^\pm\,=\,\xi^\pm_P\,-\,\xi^\pm_Q$.   Then taking the limit $\rho_{P,Q} \to \infty$, we obtain the expression for the entanglement entropy of a single interval in the conical deficit state, up to an additive constant:
\bea
S_{\rm EE}(P,Q)\,&&=\,\frac{c}{6}\,\ln\left[\frac{2e^{\rho_P+\rho_Q}}{\alpha^2}\,\sin\left(\alpha\,\frac{\Delta \xi^+}{2}\right)\,\sin\left(\alpha\,\frac{\Delta \xi^-}{2}\right)\right]\,,\label{EEconical}\\\nonumber\\\nonumber
\alpha\,&&\equiv\,\sqrt{1-\tfrac{24 \Delta_{\cal O}}{c}}\,.
\eea
This is the expected result for the single-interval covariant entanglement entropy in the conical deficit state.

\paragraph{Infalling particle:}  The backreacted geometry associated to an infalling particle of mass $m$ in the BTZ background with temperature $\beta^{-1}$ is obtained by a  coordinate transformation and boost on the conical deficit state. When $\delta =0$, the map simply transforms global ${\rm AdS}_3$ to the BTZ black hole. The explicit form  \eqref{fullshock} of the coordinate transformation in \cite{Caputa1}
\footnote{Recently, these coordinate transformations have been applied to study evolution of entanglement in holographic bilocal quenches in ${\rm CFT}_2$  \cite{Arefeva:2017pho}.}, also includes a boost parameter $\tilde\epsilon$ that is directly related to the `width' $\sim \epsilon$ of the local quench \cite{Caputa1, paper1}. For the holographic entanglement entropy, calculated using the Wilson line prescription, we only need to know how the coordinates of the endpoints of the Wilson line on the conformal boundary transform under this map:
\bea
&& e^{\rho_{P,Q}}\, =\,\label{caputa_transf}\\\nonumber\\\nonumber
&&\frac{ \Lambda R \beta}{2\pi} \sqrt{ \sinh^2 \left( \frac{2\pi\, x_{P,Q}}{\beta} \right)+\left(\frac{\beta}{2\pi\tilde\epsilon}\,\cosh\left(\frac{2\pi t}{\beta} \right)\,-\,\sqrt{\left(\tfrac{\beta}{2\pi\tilde\epsilon}\right)^2-1}\, \cosh \left( \frac{2\pi x_{P,Q}}{\beta}\right)\right)^2}\nonumber\\\nonumber\\\nonumber
&& \tan \left(\tau_{P, Q}\right)\, =\, \frac{2\pi \tilde \epsilon}{\beta} \frac{\sinh \left(\frac{2\pi t}{\beta}\right)}{\cosh \left( \frac{2\pi\, x_{P,Q}}{\beta}\right)- \sqrt{1-\left(\frac{2\pi\tilde\epsilon}{\beta}\right)^2}\cosh\left(\frac{2\pi t}{\beta}\right)}\\\nonumber\\\nonumber
&& \tan \left(\phi_{P,Q}\right)\, = \,\frac{2\pi \tilde \epsilon}{\beta} \frac{\sinh \left(\frac{2\pi\, x_{P,Q}}{\beta}\right)}{\cosh \left( \frac{2\pi t}{\beta}\right)- \sqrt{1-\left(\frac{2\pi\tilde\epsilon}{\beta}\right)^2}\cosh\left(\frac{2\pi \,x_{P,Q}}{\beta} \right)}\,.
\eea
$\Lambda$ is the location of the AdS boundary which provides the UV cutoff in the CFT. 
The spatial coordinates of the endpoints of the Wilson loop in the shockwave background must be identified with endpoints of the entangling interval in the quenched state, $x_P=\ell_1$ and $x_Q=\ell_2$. The parameter $\tilde\epsilon$ is related to the width $\epsilon$ of the quench via
\be
\frac{2\pi\tilde\epsilon}{\beta}\,=\,\sin \frac{2\pi\epsilon}{\beta}\,.
\ee
The coordinate transformations above are simply conformal transformations on the boundary acting as,
\be
e^{i\xi^\pm}\,=\,e^{2\pi i\epsilon/\beta}\,\frac{\sinh\tfrac{\pi}{\beta}(t\,\pm\,x-i\epsilon)}{\sinh\tfrac{\pi}{\beta}(t\,\pm\,x+i\epsilon)}\,.
\ee
These extend into the bulk as diffeomorphisms which act as the diagonal subgroup of 
${\rm SL}(2,{\mathbb R})\times {\rm SL}(2,{\mathbb R})$ gauge transformation 
on the Chern-Simons connections. The diagonal subgroup leaves the Wilson line \eqref{wlee} invariant.
The transformation \eqref{caputa_transf} is explored in more detail in appendix \ref{appendix:gtbtz}.

\begin{figure}[h]
\begin{center}
\includegraphics[width=2.3in]{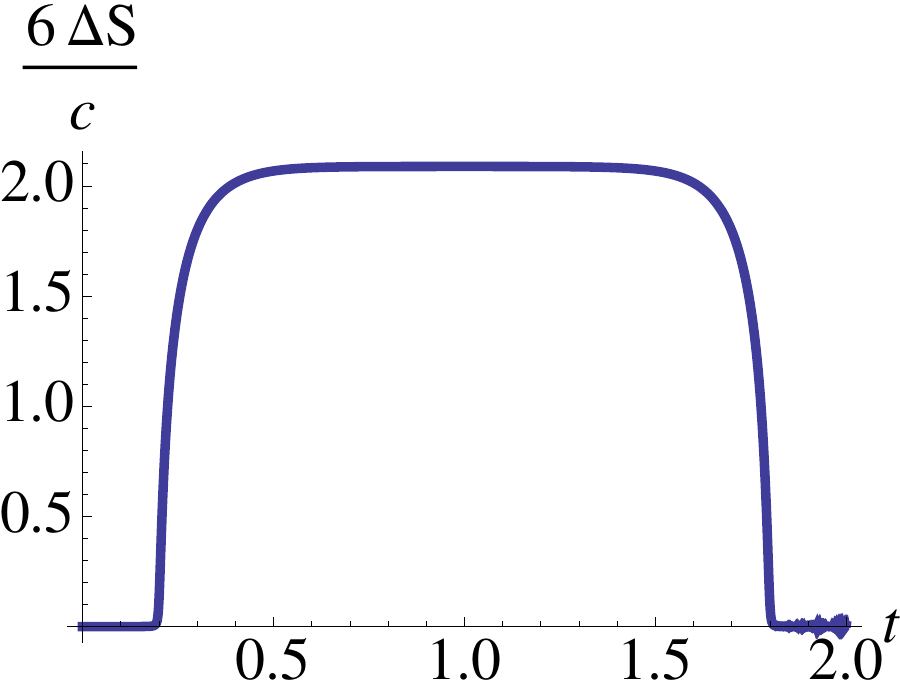}
\hspace{0.9in}\includegraphics[width=2.3in]{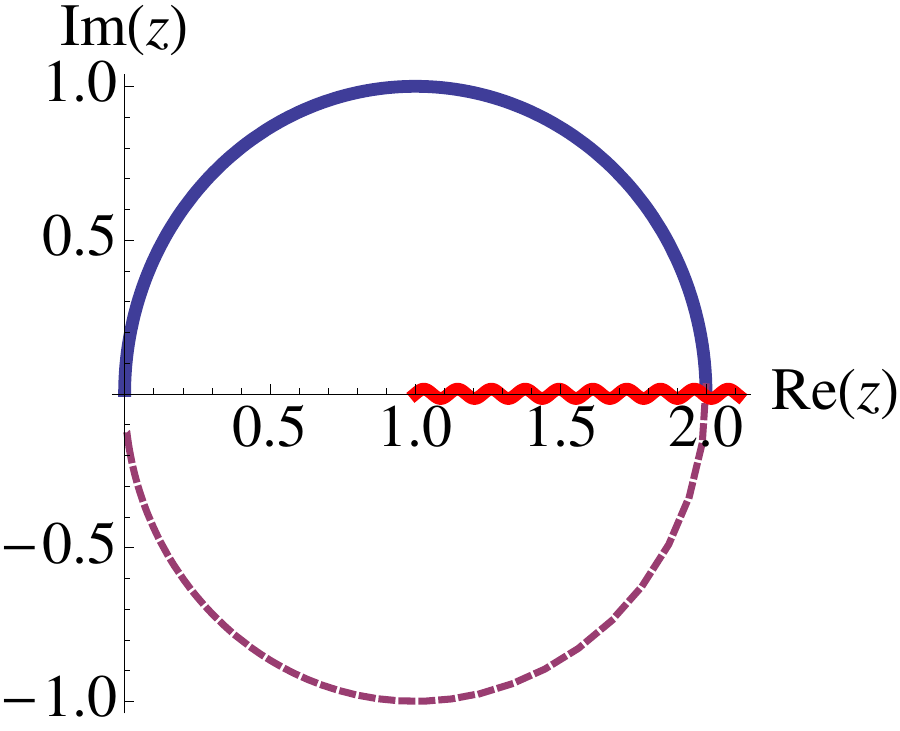}
\end{center}
\caption{\small {\bf Left:}The change in the entanglement entropy as a function of time at $\beta=0.5$ with $\ell_1=0.2$, $\ell_2=1.8$, following a quench by an operator with $\Delta_{\cal O}/c\,=\,0.01$ and quench width $\epsilon\,=\,0.005$. {\bf Right:} The cross-ratio $z$ traverses clockwise around the branch point at $z=1$, crossing the branch cut, and moving to the next sheet when the excitation enters the light cone of the endpoint at $x=\ell_1$.}
\label{fig:sl2ee}
\end{figure}

Substituting the transformed coordinates into the expression \eqref{EEconical} we obtain the time dependence of the single interval entanglement entropy in the presence of a local quench. An important point here is that the global coordinates $(\tau_{P,Q},\, \phi_{P,Q})$ are multivalued functions of $(t,\, x_{P,Q})$, and it is necessary to identify and choose the branches which are appropriate for describing early times ($t< \ell_1$), intermediate times $(\ell_1< t< \ell_2)$ and late times $t> \ell_2$, following the local quench. Figure \ref{fig:sl2ee} shows the behaviour of the change in the entanglement entropy as a function of time.

The entanglement entropy \eqref{EEconical} is best expressed in terms of the conformal cross-ratios $z$ and $\bar z$ involving the locations of the two heavy $\left({\cal O}^\dagger,\,{\cal O}\right)$ and two light $({\cal T}, \widetilde{\cal T})$ operators:
\be
z\,\equiv\,\frac{(z_2-z_3)(z_1-z_4)}{(z_2-z_1)(z_3-z_4)}\,,\qquad\qquad z_i\,\equiv\, e^{\frac{2\pi}{\beta} x_i}\,, \qquad i=1,2,3,4\,,\label{crossratio}
\ee 
where the insertion points $\{x_i\}$ are defined in \eqref{hhllpoints}. The anti-holomorphic cross-ratio
%The twist operators that 
%\be
%z\,=\,\frac{\sinh\left[\tfrac{\pi}{\beta}(x_1\,-\,x_2)\right]\,\sinh\left[\tfrac{\pi}{\beta}(x_4\,- \,x_3)\right]}{\sinh\left[\tfrac{\pi}{\beta}(x_1\,-\,x_3)\right]\,\sinh\left[\tfrac{\pi}{\beta}(x_4\,-\,x_2)\right]}\,,
%\ee
$\bar z$ is defined in the same way using ``barred" or anti-holomorphic coordinates on the thermal cylinder.  In terms of these cross-ratios we then have,
\bea
W_{\rm fund}&&\,=\,\,\exp\left(\frac{6}{c}S_{\rm EE}(PQ)\right)\\\nonumber\\\nonumber
&&\,=\,\frac{8 (R\Lambda)^2\beta^2}{\pi^2 \alpha^2} \,\sinh^2\left[\tfrac{\pi}{\beta}(\ell_2-\ell_1)\right]\,\frac{\left(1\,-\,(1-z)^\alpha\right)\,\left(1\,-\,(1-\bar z)^{\alpha}\right)}{(1-z)^{\frac{\alpha-1}{2}}\,(1-\bar z)^{\frac{\alpha-1}{2}}\,z\,\bar z}\,.
\eea
In obtaining the final form of this expression, an overall factor $\sim\exp(2\pi\,(\ell_1+\ell_2)/\beta)$ has been accounted for by the covariant tensor transformation law for the twist field correlators under the exponential map from the plane to the (thermal) cylinder. Holographically, this is understood as a rescaling of the location of the UV-cutoff/boundary to which the Wilson line is anchored. The ${\rm SL}(2,{\mathbb R})$ Wilson line computes the length of the geodesic joining the two endpoints on the boundary, in the background generated by the infalling massive particle. Within the standard AdS/CFT dictionary, this provides  the four-point correlator 
\be
\left(W_{\rm fund}\right)^{-2 \Delta_{\cal T}}\,=\,\frac{\langle {\cal O}^\dagger(x_1,\bar x_1)\,{\cal T}(x_2, \bar x_2)\,
\widetilde {\cal T}(x_3, \bar x_3)\,{\cal O}(x_4, \bar x_4) \rangle}
{\langle {\cal O}^\dagger(x_1,\bar x_1)\,{\cal O}(x_4, \bar x_4) \rangle} \,.
\ee
Here $\Delta_\tau$ is the scaling dimension of the operator ${\cal T}$. For twist fields computing the entanglement entropy of the interval $PQ$  we need to take 
${\cal T}\,=\, \frac{c}{24}\left(n-\frac{1}{n}\right)$ in the limit $n\to 1$.

\paragraph{Out-of-time ordering:} The key feature of this expression which is responsible for  nontrivial time dependence in the entanglement entropy when the local perturbation enters the interval $PQ$, is the presence of a branch point at $z=1$. In particular, when the excitation enters the lightcone of one of the endpoints of the interval, the cross-ratio traverses clockwise around the branch-point so that 
\be
(1-z)\,\to (1-z)\,e^{-2\pi i}\,.
\ee
For $\epsilon \ll \beta$, this is shown in figure \ref{fig:sl2ee}. This does not affect $\bar z$. In fact, as explained in \cite{Perlmutter:2016pkf}, this rotation  of the cross-ratio $z$ yields precisely the out-of-time ordered (OTO) configuration of the four operators. Explicitly, the cross-ratio $z$, as a function of (real) time $t$ is
\be
z\,=\,\frac{i\sin\left(\tfrac{2\pi\epsilon}{\beta}\right)\sinh\frac{\pi}{\beta}(\ell_2-\ell_1)}{\sinh\frac{\pi}{\beta}(\ell_1-t+i\epsilon)\,\sinh\frac{\pi}{\beta}(\ell_2-t-i\epsilon)}\,.\label{zdef}
\ee 
Therefore, for generic $t$, the cross-ratio $z$ is $O(\epsilon)$ in the limit of small $\epsilon$. It is useful to define the quantity, 
\be
{\cal Z}_{\ell_1,\ell_2}(t)\,\equiv\,
\frac{\sinh\frac{\pi}{\beta}(\ell_2-\ell_1)}{\sinh\frac{\pi}{\beta}(t-\ell_1)\,\sinh\frac{\pi}{\beta}(\ell_2-t)}\,.
\ee 
This function will appear repeatedly at various points below, when we consider the limit of small $\epsilon$.
We can now calculate the change in the Wilson line correlator  or equivalently,  the change in the entanglement entropy 
$\Delta S_{\rm EE}$.  We will compute this in a double scaling limit such that,
\be
\frac{\Delta_{\cal O}}{c}\,=\, \frac{E}{\pi c}\,\epsilon\, \ll 1\,,\qquad\qquad 
\epsilon\ll \beta\,,
\ee
with $E/\pi c$ fixed in the large-$c$ limit. Using this scaling, we  perform an expansion in powers of $\epsilon$ and find:
\bea
\frac{6}{c}\,\Delta S_{\rm EE}\big|_{\ell_2>t>\ell_1}\,=&&
\ln W_{\rm fund}\,-\,
\left.\ln W_{\rm fund}\right|_{t =0}\,
\label{smallz} \\\nonumber
= && \ln\left(1\,+\,12\left(\frac{\beta E}{\pi\, c}\right)\,{\cal Z}^{-1}_{\ell_1,\ell_2}(t)\right)\,+\,O(\epsilon)\,.\nonumber
\eea
The height of the jump in $\Delta S_{\rm EE}$ is thus positive and in the limit of large interval length $(\ell_2\to\infty)$, it asymptotes to the late time value, 
\be
\Delta S_{\rm EE}\Big|_{\ell_2\to\infty,\,t\gg \ell_1}\,=\,\frac{c}{6}\ln\left(1+\frac{6\beta}{\pi c}E\right)\,=\,\frac{c}{6}\ln\left(1\,+\,\frac{2E}{S_\beta}\right)\,. \label{sl2max}
\ee
The jump is positive definite, and its magnitude is determined by the ratio of the injected energy $E$ to the thermal entropy density $S_\beta\,=\,\pi c/3\beta$. The size of the jump displayed in figure \ref{fig:sl2ee} is in agreement with the value obtained above.
In fact, precisely the same ratio enters in the scrambling time \cite{Shenker:2013pqa, Maldacena:2015waa}. As explained below, this is not a coincidence. 
 
We may consider a particular limit of the expression \eqref{smallz} which brings us to the Regge limit of the OTO correlator discussed in \cite{Perlmutter:2016pkf}. In this context we note the following two points: (i) The rotation of the cross ratio about the branch point at $z=1$ yields the out-of-time ordering of the operators in question. (ii) In addition, once within the regime $\ell_2 > t >\ell_1$ we also naturally have $z \ll 1$ since $z\,\sim  \,O(\epsilon)$. Therefore this branch of the correlator and its small $\epsilon$ expansion has an overlap with the Regge limit of large times discussed in \cite{Perlmutter:2016pkf}. 

In the regime of intermediate times $\ell_2 > t>\ell_1$, the cross ratio $z$ is small, and pure negative imaginary,
\be
z\approx -\frac{2\pi i \epsilon}{\beta} {\cal Z}_{\ell_1,\ell_2}(t).
\ee 
In order to see the onset of chaos we require the Regge limit $|z| \ll 1$, and  $t\gg \ell_2$ on the second sheet. For simplicity, we also take $\ell_2 \gg \ell_1$, but this is not really necessary. In this limit,
\be
z\approx \frac{4\pi i \epsilon}{\beta} e^{-\frac{2\pi}{\beta}(t-\ell_2)}\,.
\ee
The continuation from ${\rm Im}(z)<0$ to ${\rm Im }(z)>0$ appears to be a simple phase rotation of $z$. However, this is not completely straightforward when $z$ is viewed as a function of time. To get to the chaos regime on the second sheet we cannot simply take $t$ to be large, since we must avoid the singularity at $t=\ell_2$. In appendix \ref{app:zrot} it is shown how this can be achieved by adding an imaginary part to $t$ so that $t \to t+ i\epsilon_1$   where $\epsilon_1$ is parametrically larger than $\epsilon$. We  may then take   the limit $|t| \gg \ell_2$,  and subsequently  $\ell_2 \gg \ell_1$, whilst remaining on the second sheet, yielding
%\footnote{Viewed as a function of time $t$, this operation requires some clarification as there is an issue of the order of limits. We have taken the limit of small $\epsilon$ for the correlator in the second sheet and then taken $t$  to large values. This can be understood more carefully as an analytic continuation of the function whilst remaining on the second sheet. On the other hand, when viewed as a small $z$ expansion on the second sheet, there is no ambiguity because the Regge limit for chaos has an overlap with the small $\epsilon$ limit (both correspond to $|z|\ll 1$ in the second sheet).},
\be
W_{\rm fund}\Big|_{\rm OTO}\,\to\,\frac{8 (R \Lambda)^2 \beta^2}{\pi^2}\,\sinh^2\tfrac{\pi}{\beta}\left(\ell_2-\ell_1\right)\,\left(1\,-\,\frac{6\beta E}{\pi\,c}\,e^{\frac{2\pi}{\beta}(t\,+\,i\epsilon_1-\ell_2)}\right)\,.
\ee 
We thus identify the Lyapunov exponent $\lambda_L \,=\,\frac{2\pi}{\beta}$ which controls the late time, exponential departure of the OTO correlator from its constant value\footnote{Note that correlators of operators which are computed holographically by geodesics or Wilson lines, are obtained from the geodesic actions $S_{\rm geo}=\,\ln W(P,Q)$ by exponentiating the latter  $\sim e^{- (2\Delta)\,S_{\rm geo}(PQ) }$
 where ${\Delta}$ is the operator dimension. The analogous expression for the Wilson line is $\sim W(P,Q)^{-2 \Delta} $. Therefore the exponential growth in time of the Wilson line implies a decaying correlation function.}. Note that this {\em does not} dictate the behaviour of the single interval entanglement entropy following the local quench. The time evolution of the physical $\Delta S_{\rm EE}$ for times $t> (\ell_1+\ell_2)/2$ is obtained by  reversing  the rotation of $z$ around the branch point at $z=1$ as the excitation exits the interval and we obtain the form depicted in figure \ref{fig:sl2ee}. What we have described above is a particular analytic continuation of the single-sided Wilson line correlator  to the second sheet,  which yields the chaotic behaviour of the OTO correlator in the late time limit.  When $\tilde \epsilon\,=\,\beta/2$, we obtain the expression for the two-sided correlator in the thermofield double state, and our expressions may then be matched with the corresponding ones in \cite{Shenker:2013pqa, Perlmutter:2016pkf}. 
The scrambling time $t_*$ also follows from the analytically continued single-sided Wilson line. It is given by the time scale at which the exponentially growing term becomes of order one,  characterising the decay time of the OTO correlation functions,
\be
t_*\,=\,\ell_2\,+\,\frac{\beta}{2\pi}\,\ln\left(\frac{ S_\beta}{2E}\right)\,.
\ee

\section{ Quench with spin three charge}
\label{sec:sl3ee}
In ${\rm SL}(N, {\mathbb R})\times {\rm SL}(N, \mathbb R)$ Chern-Simons theory with $N> 2$ the Wilson line which computes entanglement entropies is defined in a representation ${\cal R}$, whose dimension rises exponentially with $N$ \cite{deBjottar}, 
\be
{\rm dim}[{\cal R}]\,=\,2^{N(N-1)/2}\,.
\ee
 For the case $N=3$, the appropriate representation is the 8 dimensional or adjoint representation.  

\paragraph{Spin-three conical deficit:}In the ${\rm SL}(3, {\mathbb R})$ theory the conical deficit state can be endowed with spin three charge, and we can then analyse its effect on the single interval entanglement entropy and the OTO correlator. We will do this first in the canonical ensemble with fixed spin-three charge. The constant flat connections relevant for a  charged, conical deficit state are:
\bea
&&a\,=\,a_+\,d\xi^+\,,\qquad\qquad \bar a\,=\,a_-\,d\xi^-\,,\\\nonumber\\\nonumber
&& a\,=\,\left(L_1\,-\,\frac{\pi{\cal L} }{2k_{\rm cs}}\, L_{-1}\,-\,\frac{\pi {\cal W}}{8 k_{\rm cs}}\,W_{-2}\right)\,d\xi^+\,,\\\nonumber\\\nonumber
&& \bar a\,=\,-\left(L_{-1}\,-\,\frac{\pi \bar{\cal L}}{2k_{\rm cs}}\, L_{1}\,-\,\frac{\pi \overline{\cal W}}{8 k_{\rm cs}}\,W_{2}\right)\,d\xi^-\,.
\eea
%  where the mass and spin-three charges can be expressed in notation that is conventional  in the higher spin literature \cite{Gutperle:2011kf},
%  \be
%  M\,=\,\frac{2\pi{\cal L} }{k_{\rm cs}}\,\,,\qquad  \overline M\,=\,\frac{2\pi \bar{\cal L}}{k_{\rm cs}}\,\,,\qquad Q_3\,=\,\frac{\pi {\cal W}}{k_{\rm cs}}\,\,,\qquad \overline Q_3\,=\,\frac{\pi \overline{\cal W}}{k_{\rm cs}}\,\,.
% \ee
Here ${\cal W}$ and $\overline{\cal W}$ are the spin three charges in each sector, and $W_{\pm 2}$ are generators of the ${\rm SL}(3, {\mathbb R})$ (appendix \ref{app:sl3}).
  We will restrict ourselves to the  so called non-rotating background with ${\cal L}\,=\,\bar{\cal L}$ and ${\cal W}\,=\,-\,\overline{\cal W}$. As in the pure gravity case, the conical deficit states have negative energy, so that ${\cal L}<0$ and,
  \be
 \frac{\pi{\cal L} }{2k_{\rm cs}}\,=\,-\,\frac{1}{4}\alpha^2\,=\,-\,\frac{1}{4}\left(1\,-\,\frac{24\Delta_{\cal O}}{c}\right)\,, \qquad c\,=\, 24\, k_{\rm cs}\,.
  \ee
Here ${\Delta}_{\cal O}$ is the dimension of the heavy operator which now also carries spin three charge ${\cal W}$.
  
  The computation of the Wilson line requires us to exponentiate the constant connections. This is best done in the diagonal basis and subsequently relating to the original basis via a similarity transformation. The eigenvalues of $a^\pm$ in the defining representation are given by  the roots $\{\nu_i\}$ of the cubic equation,
 \be
 \nu^3\,+\,\alpha^2\,\nu\,+\,\frac{\pi{\cal W}}{k_{\rm cs}}\,=\,0\,.\label{cubic}
 \ee
The form of the cubic immediately implies that the three roots must satisfy the constraints, 
\bea
\nu_1\,+\nu_2\,+\,\nu_3\,=\,0\,,\qquad \nu_1\nu_2\,+\,\nu_2\nu_3\,+\,\nu_3\nu_1\,=\,\alpha^2\,,\qquad \nu_1\nu_2\nu_3\,=\,-\,\frac{\pi{\cal W}}{k_{\rm cs}}\,.\label{product}
\eea
The eigenvalues for the matrices in the adjoint representation are then $\{\pm \lambda_1,\, \pm \lambda_2,\, \pm \lambda_3,$ $\,0,\,0\}$ where we have defined
\bea
&&\lambda_1\,=\,\nu_1\,-\nu_2\,,\qquad \lambda_2
\,=\,\nu_2\,-\,\nu_3\,,\qquad \lambda_3\,=\, \nu_3\,-\,\nu_1\,.
\eea
%The definitions automatically ensure that
%\be
%\lambda_1\,+\,\lambda_2\,+\,\lambda_3\,=\,0\,,\qquad
%\bar\lambda_1\,+\,\bar\lambda_2\,+\,\bar\lambda_3\,=\,0\,.
%\ee
We can then compute (see appendix \ref{app:wl}) the  ${\rm SL}(3,{\mathbb R})\times{\rm SL}(3,{\mathbb R}) $ Wilson line in the adjoint representation, and we find,
\bea
\lim_{\rho\to\infty} &&W_{\rm Ad}(P,Q)\,=\,\label{WLadj}\\\nonumber\\\nonumber
&&\frac{256\,e^{4(\rho_P + \rho_Q)}}{\left(\lambda_1\lambda_2\lambda_3\right)^2}\,\left[\sum_{i=1}^3{\lambda_i^{-1}\,\sinh^2\left(\frac{\lambda_i}{2}\,\Delta \xi^+\right)}\right]\left[\sum_{i=1}^3\lambda_i^{-1}\,\sinh^2\left(\frac{\lambda_i}{2}\,\Delta \xi^-\right)\right]\,.
\eea
 \subsection{Wilson line in charged shockwave background}
 We expect that the effect of the local quench by the charged operator ${\cal O}$ should follow from the corresponding boosted, transformed conical deficit state discussed above. We note that in a higher spin theory the spacetime metric is not gauge-invariant and  therefore the operations on the bulk geometry do not have an obvious invariant meaning. It would be interesting to understand the generalisation to the higher spin situation (in Chern-Simons language) of the coordinate transformations that map the conical deficit in ${\rm AdS}_3$ to the infalling shockwave geometry.
  Nevertheless, given that the corresponding transformations are defined as before in the boundary CFT, we use the transformations \eqref{caputa_transf} to evaluate the  higher-spin Wilson line correlator in the charged quenched state. The ${\rm SL}(3,{\mathbb R})$ Wilson line  can be rewritten in terms of the cross ratio $z$ as  in the  pure gravity theory:
 \bea
W_{\rm Ad}(P,Q)&&\,=\,\frac{2^{14}(R\Lambda)^8\beta^8}{\pi^8}\sinh^8\tfrac{\pi}{\beta}(\ell_2-\ell_1)\times
\\\nonumber
(\lambda_1\lambda_2\lambda_3)^{-2}&&(z\,\bar z)^{-4}\left[\sum_{j=1}^3\frac{1}{\lambda_j}\left(\frac{(1-z)^{i\lambda_j}-1}{(1-z)^{i\lambda_j/2-1}}\right)^2\right]\left[\sum_{j=1}^3\frac{1}{\lambda_j}\left(\frac{(1-\bar z)^{i\lambda_j}-1}{(1-\bar z)^{i\lambda_j/2-1}}\right)^2\right]\,.
\eea
The roots $\nu_i$ of the cubic \eqref{cubic}  are complicated functions of the mass $\alpha$ and the spin-three charge ${\cal W}$. However, as we saw in the pure gravity situation, we will be interested in a double scaling limit, where the deficit angle and spin-three charge are both small in the limit of small $\epsilon$:
\be
\frac{\Delta_{\cal O}}{c}\,=\,\frac{E}{\pi c}\,\epsilon\,,\qquad\qquad {\cal W}\,=\,\epsilon^2\, \frac{4q}{\pi^2}\,,\qquad \epsilon \ll \beta\,.
\ee
We view this as a double scaling limit because in the holographic gravity description which applies in the limit $c\to\infty$, we are also keeping $\Delta_{\cal O}/c$ and $q/c$ fixed. With this scaling, the roots of the cubic \eqref{cubic} are,
\be
\nu_{1,2}\,\simeq \,\pm i\left[1-\frac{12 E}{\pi c}\epsilon-\frac{72 E^2}{\pi^2 c^2}\epsilon^2 \right]\,+\,\frac{48}{\pi} \frac{q}{c}\,\epsilon^2\,,\qquad\qquad\nu_3\,\simeq\, -\frac{96}{\pi} \frac{q}{c}\,\epsilon^2
\ee
When the charged excitation generated by operator ${\cal O}$ enters the interval $PQ$, for intermediate times $\ell_2 > t > \ell_1$, the cross-ratio  traverses clockwise around the branch point of the Wilson line at $z=1$, bringing us into the out-of-time-ordered configuration,
\be
(1-z)\,\to\,(1-z)\,e^{-2\pi i}\,.
\ee 
Following this, we expand the Wilson line in the small $\epsilon$ limit. As explained previously, $z\sim O(\epsilon)$. Therefore, at the leading order $\sim {O}(\epsilon^0)$, the adjoint Wilson line is,
\bea
W_{\rm Ad}(P, Q)&&\simeq\,\label{wadz}\\\nonumber
&& \frac{2^{12}(R\Lambda)^8\beta^8}{\pi^8}\sinh^8\tfrac{\pi}{\beta}(\ell_2-\ell_1)\left[\left(1\,-\,\frac{24iE}{c}\frac{\epsilon}{z}\right)^4\,-\,\frac{\,q^2}{c^2}\left(\frac{24\epsilon}{ z}\right)^4\right]\,+{O}(\epsilon)\,.
\eea
Substituting the expression for $z$ as a function of real time \eqref{zdef} in the interval $\ell_2 > t> \ell_1 $ we find the change in entanglement entropy of the interval $PQ$, following the spin-three local quench (note that $c=24 k_{\rm cs}$ for the ${\rm SL}(3, \mathbb R)\times {\rm SL}(3, \mathbb R)$ theory):
\bea
\Delta S_{\rm EE}(PQ)\,=&&\frac{c}{24}\left(\ln W_{\rm Ad}\,-\,\ln\left.W_{\rm Ad}\right|_{t=0}\right)\\\nonumber\\\nonumber
&&=\,\frac{c}{24}\,\ln\left[\left(1\,+\,\frac{12\beta E}{\pi c}{\cal Z}^{-1}_{\ell_1,\ell_2}(t)\right)^4\,-\,\frac{q^2}{c^2}\left(\frac{12\beta}{\pi} \,{\cal Z}^{-1}_{\ell_1,\ell_2}(t)\right)^4\right]\,+\,O(\epsilon).
\eea
The result reveals some important features.  For large intervals $\ell_2 \gg \ell_1$, the change in the entanglement entropy following the entry of the perturbation into the interval saturates to a maximal value given by,
\be
\Delta S_{\rm EE}\Big|_{\ell_2\to\infty, \,t\gg \ell_1}\,=\,
\frac{c}{24}\,\ln\left[\left(1+\frac{6 \beta E}{\pi c}\right)^4\,-\,\frac{q^2}{c^2}\left(\frac{6\beta}{\pi}\right)^4\right]\,.\label{sl3max}
\ee
Since $q$ is a dimension two charge, obtained by integrating a dimension three current, all terms within the argument of the logarithm are dimensionless.   
Further, we are taking $E/c$ and $q/c$ to be fixed in the large-$c$, classical gravity limit. For vanishing $q$, 
eq.\eqref{sl3max} matches the pure gravity value \eqref{sl2max}.  
 With $q\neq 0$, the argument of the logarithm is no longer positive definite. Requiring that the result for $\Delta S_{\rm EE}$ be  physically sensible (i.e. avoiding divergent or complex values), we obtain a restriction on the spin-three charge that the perturbing operator/state can carry:
\be
\sqrt {\frac{|q|}{c}}\,<\,\frac{E}{c}\,+\,\frac{S_\beta }{2 c}\,,\qquad\quad S_\beta\,=\,\frac{\pi c}{3\beta}\,.
\ee 
Multiplying through by the width $\epsilon$ of the quench and taking the limit $\epsilon/\beta \to 0$, this yields the bound,
\be
\sqrt {\frac{|{\cal W}|}{c}}\,<\,\frac{2\Delta_{\cal O}}{c}\,,\label{critical}
\ee
where it is understood that the both quantities $\Delta_{\cal O}/c$ and ${\cal W}/c$ are fixed and small ($O(\epsilon)$ and $O(\epsilon^2)$ respectively) in the large $c$ limit. There is no independent reason to expect generic charged operators in a  large-$c$ CFT  with ${\cal W}_3$ symmetry to respect this requirement. If such  bound is not realised, one would have to conclude that corresponding  theories are unphysical.

\begin{figure}[h]
\begin{center}
\includegraphics[width=1.8in]{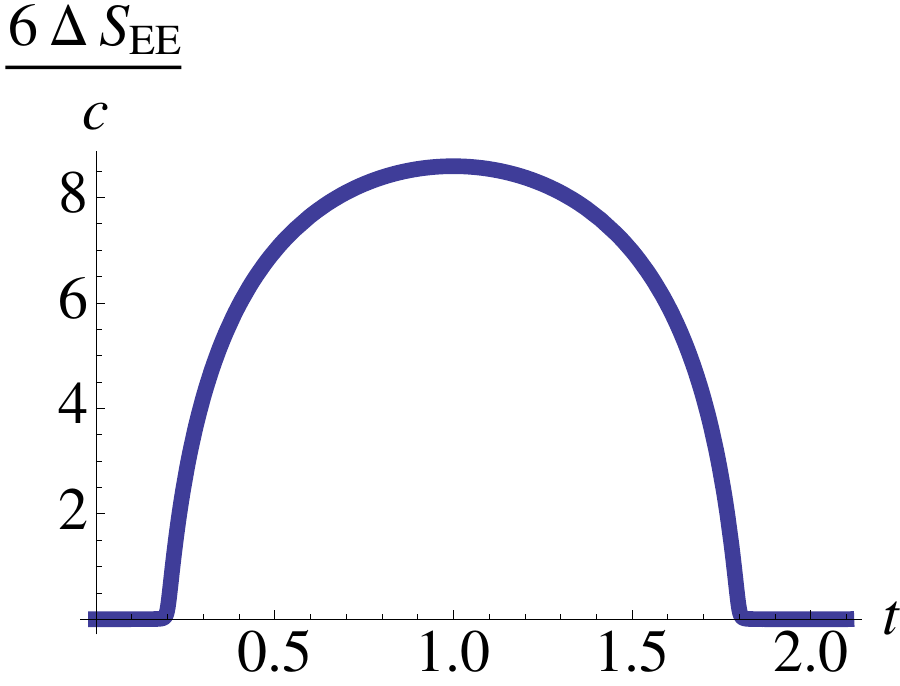}\hspace{0.2in}\includegraphics[width=1.8in]{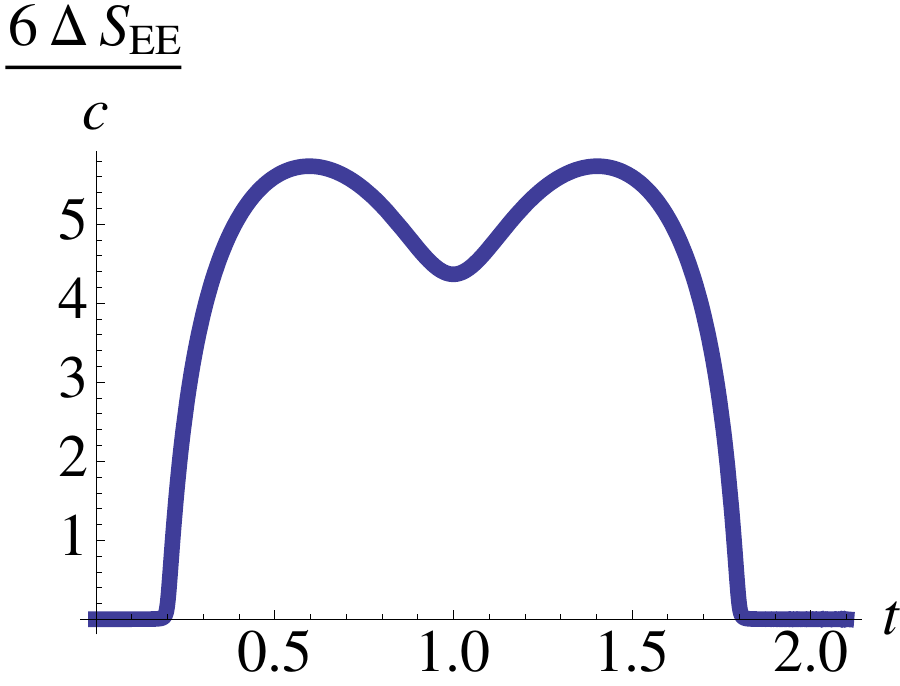}\hspace{0.2in}\includegraphics[width=1.8in]{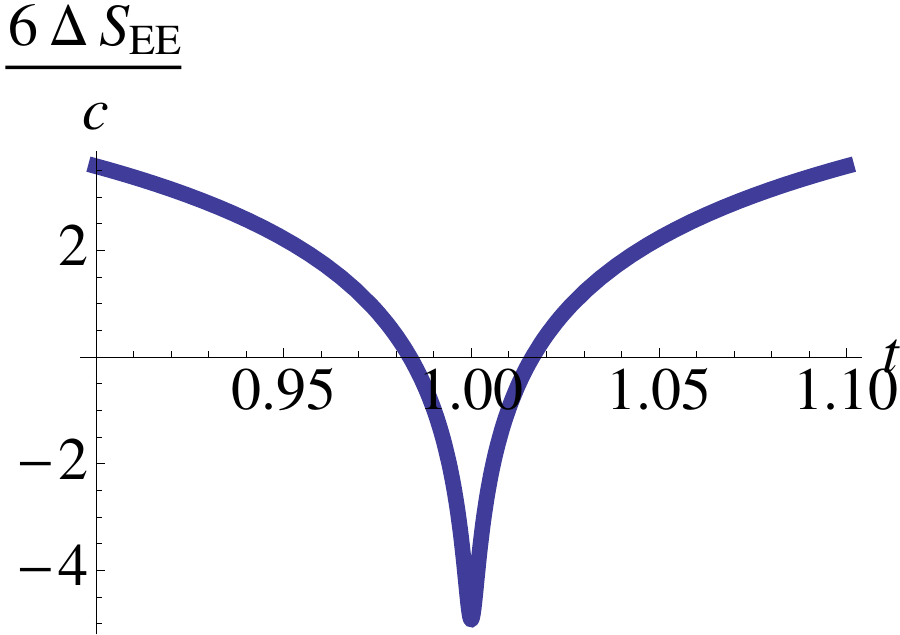}
\end{center}
\caption{\small{The change in holographic entanglement entropy following a local quench by an operator carrying spin-three charge in ${\rm SL}(3,{\mathbb R})\times {\rm SL}(3,{\mathbb R})$ Chern-Simons theory. The quench width is $\epsilon=0.008$, $\beta=5.5$, $\ell_1=0.2$, $\ell_2=1.8$, and $\Delta_{\cal O}/c=0.5$}. The figure on the left has $q/c=0$, the central one corresponds to $q/c=2.884$ whilst the rightmost one depicts the onset of a divergence at a critical value $q/c \approx 2.904$.}
\label{spin3ee}
\end{figure}

Figure \ref{spin3ee} shows the behaviour of the change in the single interval entanglement entropy for a local quench carrying spin-three charge. The introduction of the spin-three charge has the effect of decreasing the height of $\Delta S_{\rm EE}$ for a given energy $E$, eventually driving it negative and unbounded from below at a critical value of $q$.

As in the pure gravity case described in section \ref{holomorphic_pres_review}, we can take the Wilson line correlator in the second sheet and make contact with the Regge limit of small $z$ and late times, by appropriate analytic continuation. We find,
\bea
W_{\rm Ad}\Big|_{\rm OTO}\,\to\,\frac{2^{12}(R\Lambda)^8\beta^8}{\pi^8}\,&&\sinh^8\tfrac{\pi}{\beta}(\ell_2-\ell_1)\,\\\nonumber
&&\left[\left(1\,-\,\frac{2E}{S_\beta}\,e^{\frac{2\pi}{\beta}(t\,+\,i\epsilon_1-\ell_2)}\right)^4\,-\,\left(\frac{36 q\beta^2}{\pi^2c}\,e^{\frac{4\pi}{\beta}(t\,+\,i\epsilon_1\,-\,\ell_2)}\right)^2\right]\,.
\eea
Comparing with corresponding expressions in \cite{Perlmutter:2016pkf}, we identify the spin-three Lyapunov exponent as $\lambda_L^{(3)}\,=\,\frac{4\pi}{\beta}$. The expression for the two-sided correlator in the thermofield double state follows upon setting  $\epsilon_1\,=\,\frac{\beta}{2}$. This phase rotation changes the sign of the coefficient of $E$, but not that of $q^2$, and the spin-three charge is then the source of singular behaviour of the corresponding OTO correlator, causing it to diverge at some finite late time, as also argued in \cite{Perlmutter:2016pkf}.
For generic $\epsilon_1$, if the spin-three charge $q$ is dominant compared to $E$, we obtain a scrambling time given by 
\be
t_*\,\simeq\, \ell_2\,+\,\frac{\beta}{4\pi}\,\ln\left(\frac{\pi^2 c}{36|q|\beta^2}\right)\,,
\label{spin3scrambling}
\ee
when the OTO correlator becomes vanishingly small.

\subsection{Mutual information} \label{section_charge_MI}
The determination of scrambling time can be performed elegantly in a large-$c$ CFT within the holographic setup by computing the mutual information of two entangled subsystems $A$ and $B$ \cite{Caputa2}. Specifically, we may take the two intervals to reside in the two different copies of the CFT, prepared in the thermofield double state. A local perturbation of the thermofield double state by  a primary operator created at some time in the past destroys  the correlations between the two copies. This is clearly seen by calculating the mutual information between the intervals $A$ and $B$. The time scale at which the mutual information,
 \be
 I_{A:B}\,=\,S_{\rm EE}(A)\,+\,S_{\rm EE}(B)\,-\,S_{\rm EE}(A\cup B)\,,
 \ee
  vanishes was calculated and identified with the scrambling time in \cite{Caputa2}. 

In this section we will consider the infalling massive particle with spin-three charge, which starts its motion close to the left boundary of the thermofield double obtained by Kruskal extension of the BTZ black hole. As in the previous sections, the black hole itself does not carry spin-three charge and therefore it should be viewed as the standard BTZ black hole embedded in the ${\rm SL}(3,{\mathbb R})\times {\rm SL}(3,\mathbb R)$ Chern-Simons framework.  The infalling particle background is inferred by the coordinate transformation and boost on the conical deficit state in global ${\rm AdS}_3$. This is done separately for the two AdS-Schwarzschild patches of the eternal BTZ geometry \cite{Caputa2}. For the spin-three charged state, our strategy is to compute the Wilson line EE for the conical deficit state and to use coordinate transformations to map the endpoints of the interval to corresponding endpoints  of the intervals on the two boundaries of the BTZ black hole.

The two intervals $L$ and $R$ are chosen to lie on the left and right boundaries of the extended BTZ geometry, respectively. The endpoints of the intervals are taken to be at  $(t_L, \ell_{1,2})$ for the interval $L$ and at $(t_R, \ell_{1,2})$ for the right interval $R$. The transformations from the conical deficit to the BTZ coordinates of the boundary on the left patch are (in the limit of small width $\epsilon$):
\bea
&& D_{L,i}\, =\, \left|\cosh \left(\tfrac{2\pi}{\beta} \ell_i \right)\,-\,\cosh \left( \tfrac{2\pi}{\beta} t_L\right)\right|\qquad\qquad e^{\rho_{L,i}}\,=\, \frac{R \Lambda \beta^2}{4\pi^2 \epsilon}\, D_{L,i}\qquad i\,=\,1,\,2\,.\nonumber\\\nonumber\\
&&\tan\left(\tau_{L,i}\right)\, =\,\frac{2\pi \epsilon}{\beta}\,\frac{ \sinh \left( \tfrac{2\pi}{\beta}t_L\right) }{D_{L,i}}\,, \qquad\qquad \tan\left(\phi_{L,i}\right)\,=\,-\frac{2\pi \epsilon}{\beta}\,\frac{ \sinh \left( \tfrac{2\pi}{\beta} \ell_i\right) }{D_{L,i}}\,.
\eea
These are just the leading small-$\epsilon$ limits of the complete transformations in eq.\eqref{caputa_transf}.
The corresponding transformations for the right boundary are,
\bea
&& D_{R,i}\, =\, \left|\cosh \left(\tfrac{2\pi}{\beta} \ell_i \right)\,+\,\cosh \left( \tfrac{2\pi}{\beta} t_R\right)\right|\qquad\qquad e^{\rho_{R,i}}\,=\, \frac{R \Lambda \beta^2}{4\pi^2 \epsilon}\, D_{R,i}\qquad i\,=\,1,\,2\,.\nonumber\\\nonumber\\
&&\tan\left(\tau_{R,i}\right)\, =\,-\frac{2\pi \epsilon}{\beta}\,\frac{ \sinh \left( \tfrac{2\pi}{\beta}t_R\right) }{D_{R,i}}\,, \qquad\qquad \tan\left(\phi_{R,i}\right)\,=\,-\frac{2\pi \epsilon}{\beta}\,\frac{ \sinh \left( \tfrac{2\pi}{\beta} \ell_i\right) }{D_{R,i}}\,. 
\eea
Note that these can be obtained from the left side transformations by the replacement $t_L\to t_R\,+\,i\beta/2$.  The mutual information,
\be
I_{L:R}(t_L,t_R)\, = \, S_{\rm EE}(L) \,+\, S_{\rm EE}(R)\, -\, S_{\rm EE}(L \cup R)\,,
\ee
is computed by ``connected" and ``disconnected" configurations of the bulk Wilson lines. The former corresponds to two bulk Wilson lines, $W_{\rm Ad}^{LR,1}$ and $W_{\rm Ad}^{LR,2}$, joining the endpoints $(t_L, \ell_i)$ of the left interval $L$ with corresponding endpoints $(t_R, \ell_i)$ of the interval $R$ on the right boundary. The disconnected contribution is simply the entanglement entropy of each individual interval, given by the Wilson lines $W_{\rm Ad}^{L}$ and $W_{\rm Ad}^R$ between the two endpoints of the interval on a given  boundary,
\bea
I_{L:R}(t_L,\,t_R)\,=\,\frac{c}{24}\left( \ln W_{\rm Ad}^R\,+\,\ln W_{\rm Ad}^L\,-\,
\ln  W_{\rm Ad}^{LR,1}\,-\, \ln W_{\rm Ad}^{LR,2}\right)\,.
\eea
%where the Wilson lines connect the following points,
%\bea
%L^L_{\gamma}: & \; (r_{L,1},\tau_{L,1},\phi_{L,1}) \; \text{to} \; (r_{L,2},\tau_{L,2},\phi_{L,2})\\
%L^R_{\gamma}: & \; (r_{R,1},\tau_{R,1},\phi_{R,1}) \; \text{to} \; (r_{R,2},\tau_{R,2},\phi_{R,2})\nonumber\\
%L^{LR}_{\gamma,1}: & \; (r_{L,1},\tau_{L,1},\phi_{L,1}) \; \text{to} \; (r_{R,1},\tau_{R,1},\phi_{R,1})\nonumber\\
%L^{LR}_{\gamma,2}: & \; (r_{L,2},\tau_{L,2},\phi_{L,2}) \; \text{to} \; (r_{R,2},\tau_{R,2},\phi_{R,2}).\nonumber
%\eea
The calculation of each Wilson line proceeds exactly as in the single-sided case analysed above, with the only difference arising in the coordinates of the endpoints.

\begin{figure}[h]
\begin{center}
\includegraphics[width=2.0in]{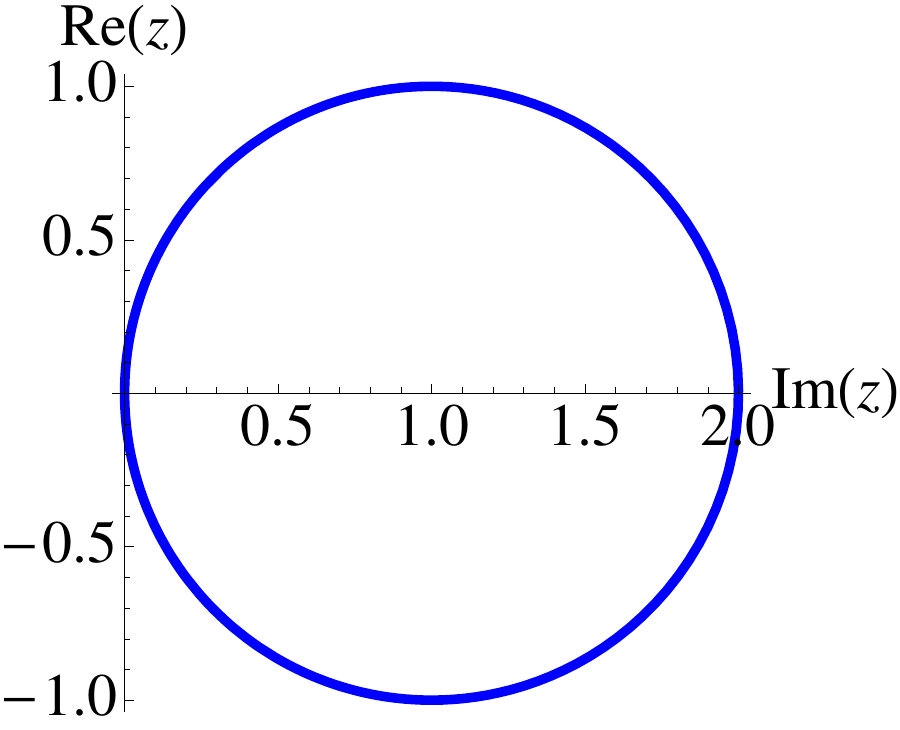}\hspace{1.0in}
\includegraphics[width=2.0in]{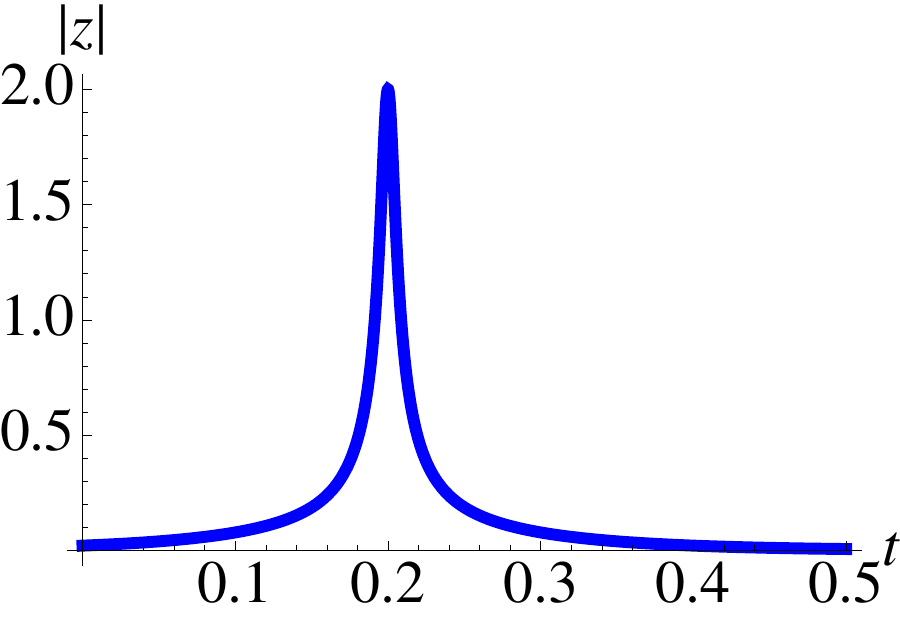}
\end{center}
\caption{\small{The cross-ratio $z=(z_{14} z_{23})/(z_{21}z_{34})$} as a function of $t$, where $z_i\,=\,\exp(2\pi x_i/\beta)$, $x_1=-i\epsilon$, $x_2=\ell_1-t$ and $x_3=\ell_1-t-i\beta/2$, $x_4=i\epsilon$. Setting $\epsilon=0.005$, $\beta=0.5$ and $\ell_1=0.2$, we see that the cross-ratio goes around $z=1$ clockwise as the excitation on the left boundary enters the interval at $t=\ell_1$.}
\label{zmutual}
\end{figure}

To calculate the scrambling time, we need the mutual information at late times, i.e. 
$t_L,t_R > \ell_{1,2}$. Thus we evaluate the Wilson lines for $t_L, t_R > \ell_{1,2}$. 
For such times, in the limit of small $\epsilon$, the EE of each interval will relax to its equilibrium value, as any local perturbation would already have exited the interval. Thus, we have
\be
W_{\rm Ad}^{L}\,=\,W_{\rm Ad}^R\,=\,\frac{2^{8}(R \Lambda )^8\beta^8}{\pi^8} \sinh ^8\tfrac{\pi}{\beta} (\ell_2\,-\,\ell_1)\,,\qquad t_{L,R} > \ell_{1,2}\,.
\ee
The connected Wilson lines can each be expressed in terms of the appropriate cross-ratio $z$ which yields, in the small-$\epsilon$ double-scaled limit, 
\bea
\label{LR_WL_charged_CD}
 W_{\rm Ad}^{LR, i}\,=\,\frac{2^{14} (R\Lambda)^8\beta^8}{ \pi^8}\, &&\cosh ^8\tfrac{\pi}{\beta}(t_L-t_R)\times\\\nonumber \\\nonumber
 &&\left[\left(1\, -\,\frac{12 E\beta}{\pi c} {\cal T}_i(t_L,t_R) \right)^4\,-\, { \frac{ q^2}{c^2}} \left(\frac{12\beta}{\pi} {\cal T}_i(t_L,t_R)\right)^4 \right]\,.
\eea
where
\be
\label{fn_def}
{\cal T}_i(t_L,t_R) \,=\, \frac{\sinh \tfrac{\pi}{\beta} (\ell_i\,-\,t_L)\,\cosh \tfrac{\pi}{\beta} (\ell_i\,-\,t_R)}{ \cosh \tfrac{\pi}{\beta} (t_L\,-\,t_R)}\,.
\ee
In terms of the cross-ratio \eqref{crossratio} the small-$\epsilon$ connected Wilson lines are given by the general formula \eqref{wadz}, with $x_1\,=\,-i\epsilon$, $x_2\,=\,\ell_i-t_{L}$, $x_3\,=\,\ell_i- t_R-i\beta/2$, $x_4\,=\,i\epsilon$.
The simplest configuration to consider is with  $t_L\,=\,t_R\,=\,t$. The resulting cross-ratio $z$ rotates clockwise around $z=1$ when the excitation on the left boundary enters the interval (see figure \ref{zmutual}) at $t=\ell_1$. 

The scrambling time $t_*$ when the mutual information vanishes then yields the condition
\bea
&&\sinh ^{16}\tfrac{\pi}{\beta}(\ell_1\,-\,\ell_2)\,=\,\\\nonumber
&&\left[\left( 1-\tfrac{12\beta E}{\pi c} {\cal T}_1(t^*) \right)^4- \frac{q^2}{c^2}\left(\tfrac{12\beta}{\pi}{\cal T}_1(t^*)\right)^4\right]\left[\left( 1-\tfrac{12\beta E}{\pi c} {\cal T}_2(t^*) \right)^4- \frac{q^2}{c^2}\left(\tfrac{12\beta}{\pi}{\cal T}_2(t^*)\right)^4\right]\,.
\eea
This can be solved easily when $t_*\gg \ell_i$ so that ${\cal T}_i\simeq -\frac{1}{4}e^{2\pi(t-\ell_i)/\beta}$ and we find the expression for the scrambling time:
\bea
\label{t*}
t_*\,=\, \frac{\ell_1+\ell_2}{2}\, +\, \frac{\beta}{\pi} \ln\left[\sinh \tfrac{\pi}{\beta} (\ell_2-\ell_1)\right]\, +\, \frac{\beta}{2\pi} \ln \left|\frac{E^4}{S_\beta^4} \,-\,\frac{9 q^2\beta^2}{\pi^2 S_\beta^2} \right|^{-\frac{1}{4}},
\eea
where $S_\beta\,=\, c\pi/(3\beta)$ is the thermal entropy density. When the  spin-three charge is vanishing we recover the known scrambling time from pure gravity. As $q$ is increased from zero, the scrambling time actually {\em increases} until it diverges at a critical value given by, $|q|/c\,=\,E^2/c^2$.  The critical value is in fact what we have already encountered above in eq.\eqref{critical}. 
Beyond this critical value the scrambling time {\em decreases}. When $E$ is negligible, the scrambling time matches what we expect for a Lyapunov exponent of $\lambda_L^{(3)}\,=\,4\pi/\beta$ (see eq.\eqref{spin3scrambling}).

 After this work was completed,  \cite{Afkhami-Jeddi:2017idc} appeared 
 which obtains constraints on the higher spin charges from unitarity and modular invariance for CFTs with $W_N$ symmetry. For the specific case of $N=3$, we can identify the spin three charge ${\cal W}$ in this paper with the corresponding quantity $q_3$ in  \cite{Afkhami-Jeddi:2017idc}. Then taking the large-$c$ limit, with $\Delta/c$ and ${\cal W}/c$ fixed, we find that our bound \eqref{critical} is consistent with the bounds obtained in  \cite{Afkhami-Jeddi:2017idc}, i.e. once the latter bound is satisfied, the bound \eqref{critical} is guaranteed to hold. This conclusion appears to depend delicately on the numerical coefficient (the factor of two on the right hand side) appearing in the inequality \eqref{critical}. Thus, unitarity which resulted in the bound of 
\cite{Afkhami-Jeddi:2017idc} ensures that there is no pathology in the behaviour of the EE.

\section{Conical defect deformed by HS chemical potential} \label{HS_CD}
The local quench above is holographically described by a particle carrying higher spin charge falling into the horizon of a bulk (uncharged) black hole. From this bulk perspective it is also possible to consider charged conical deficit states in the higher spin grand canonical ensemble i.e. with a chemical potential for higher spin charge. After computing the Wilson line in such a background we will transform the boundary endpoints so the resulting state describes an infalling charged configuration in a black hole or thermal state. From a CFT viewpoint, such a state may be viewed as the density matrix:
\be
\hat \rho_\epsilon\,=\,{\cal N}\,e^{-i Ht}\left(\sum_n e^{\,\mu_{\cal O}\beta \,{\cal W}_{{\cal O}_n}}\,{\cal O}_n(x_1,\,\bar x_1)\,e^{-\beta H}\, {\cal O}_n^\dagger(x_4,\,\bar x_4)\right)e^{i Ht}\,.
\ee 
Here ${\cal W}_{{\cal O}_n}$ represents the spin-three charge of a local operator ${\cal O}_n$ generating the local excitation. The state is therefore {\em not} in a grand ensemble for the full CFT, but only for the local quench operators. This means that the bulk black hole state in question does not carry spin-three hair and must be viewed as the BTZ black hole embedded in the ${\rm SL}(3,{\mathbb R})\times {\rm SL}(3,{\mathbb R})$ connections.
 As we will see below the results for the change in EE and scrambling time are the same as in the basic charged conical deficit example, with the value of the charge replaced by its ensemble expectation value.

\subsection{Conical deficit connections with chemical potential}
The ${\rm SL}(3,{\mathbb R})\times {\rm SL}(3,{\mathbb R})$ gauge connections for a charged state with higher spin chemical potential are \cite{Gutperle:2011kf},
\bea 
\label{HSmu_connection}
&& a\, =\, \left( L_1\, -\, \frac{\pi {\cal L}}{2k_{\rm cs}} L_{-1}\, -\, \frac{\pi {\cal W}}{8k_{\rm cs}}\, W_{-2} \right) d\xi^+\,+\\\nonumber
&&\hspace{2.0in}\, + \,\mu_{\cal O} \left( W_2 \,-\, \frac{\pi{\cal L}}{k_{\rm cs}} W_0\, +\, \frac{\pi^2{\cal L}^2}{4k^2_{\rm cs}} W_{-2}
%\right.\nonumber\\
%&&\left.\hspace{4.8in} 
\,+\, \frac{\pi {\cal W}}{k_{\rm cs}} L_{-1} \right) d\xi^- \,,\\\nonumber\\\nonumber
&&\bar{a} \,= \, - \left( L_{-1} - \frac{\pi\bar{{\cal L}}}{2k_{\rm cs}}L_1 -\frac{\pi \overline{\cal W}}{8k_{\rm cs}}\, W_2 \right) d\xi^-\\\nonumber
&&\hspace{2.0in}-\, \bar{\mu}_{\cal O} \left( W_{-2} \,-\, \frac{\pi \bar{\cal L}}{k_{\rm cs}} W_0 \,+\, \frac{\pi^2\bar{\cal L}^2}{4k^2_{\rm cs}} W_2\, + \,\frac{\pi \overline{\cal W}}{k_{\rm cs}}\, L_1 \right)d\xi^+\,.
\eea
The ${\rm SL}(3, {\mathbb R})$ generators are specified in appendix \ref{app:sl3}. 
We set $\mu_{\cal O}\,=\,-\,\bar\mu_{\cal O}$, $\overline{\cal W}\,=\,-{\cal W}$ and ${\cal L}\,=\,\bar{\cal L}$. It is well known that the higher spin chemical potential has the effect of altering the asymptotics of the spacetime metric (in radial gauge) following from the  connections above. However, this is not a problem, particularly if we need to study the system perturbatively in the chemical potential deformation. This is in fact what we will need to do in the double-scaled, small-$\epsilon$ limit. An important aspect of the perturbative corrections in $\mu_{\cal O}$ to the physical observables in the CFT is that they are only correctly described holographically by the so-called ``holomorphic" formulation of thermodynamics and of the EE/Wilson line observables \cite{deBjottar,deBoer:2013gz, universalcft}. The holomorphic prescription  to evaluate the Wilson line in the presence of higher spin chemical potential is,
\begin{align}
W_{\rm Ad}^{\rm hol}(P,Q)\, =\, \mathrm{Tr} \left[ {\cal P } \exp \left( \int_P^Q \bar{A}_-\, d\xi^-\right) {\cal P } \exp \left( \int_Q^P {A}_+ \,d\xi^+\right) \right]\,.
\end{align}
The conical deficit solution with spin-three charge is specified by  requirements on the holonomy  of the flat connections \eqref{HSmu_connection}. Analogous to the case of the  spin-three black hole of \cite{Gutperle:2011kf}, for a conical deficit state, the holonomy of the connection in the spatial ($\phi$) direction must satisfy
\be
{\rm eval}\left(\oint a_\phi\, d\phi\right)\,=\,\left(0, \pm 2\pi i\alpha \right)\,,\qquad
\alpha\,=\,\sqrt{1-\frac{\delta}{R^2}}\,.
\ee
These are algebraic conditions  which can be solved to yield  ${\cal L}$ and ${\cal W}$ as nontrivial functions of $\mu_{\cal O}$ and $\delta$. The holonomy conditions lead to two branches of solutions (a similar situation was seen for higher spin black holes in \cite{David:2012iu} and for higher spin version of global AdS \cite{Datta:2014ypa}), of  which the one smoothly connected to global ${\rm AdS}_3$ has lower free energy and is relevant for the present discussion.
In the small width limit we will only need the leading corrections in $\mu_{\cal O}$. We find:
\bea \label{def_L_W_mu_CD}
{\cal L} \,=\, -\frac{\alpha^2 \,k_{\rm cs}}{2 \pi }\left(1\,- \,20\frac{\mu_{\cal O} ^2\, \alpha^2}{3 R^2} \right)\,+\,{O}(\mu^2_{\cal O})\,,\qquad\qquad
{\cal W} \,=\, \mu_{\cal O}\,\frac{4\alpha^4 k_{\rm cs}  }{3 \pi  R}\,.
\eea
We have already seen in the fixed charge setup that we must have $\delta\sim O(\epsilon)$ and ${\cal W} \sim O(\epsilon^2)$ in small width limit. This implies $\mu_{\cal O} \sim {O}(\epsilon^2)$ in the grand canonical ensemble for the conical deficit states. In terms of CFT quantities, to leading order in the small width $\epsilon$, we have 
\be
\frac{\mu_{\cal O}}{R}\,=\,\epsilon^2 \frac{72}{\pi}\frac{q}{c}\,.\label{muqrelation}
\ee
For the holomorphic Wilson lines we only need the forms of the connection components 
$a_+$ and $\bar a_-$. Therefore, the calculation of the entanglement entropy following the local quench proceeds in exactly the same way as in the fixed charge case. One additional comment that we make in this regard is that the coordinate transformations which map the conical deficit to the infalling particle state are 
diagonal ${\rm SL}(2,{\mathbb R})\times {\rm SL}(2,{\mathbb R})$ gauge transformations embedded in ${\rm SL}(3,{\mathbb R})\times {\rm SL}(3,{\mathbb R})$,  and  therefore solutions (flat connections) remain solutions under the map.

The results of the entanglement entropy $\Delta S_{\rm EE}$, mutual information and scrambling time $t_*$ are all given by the same formulae we encountered in section \ref{sec:sl3ee}, with $q/c$ replaced by the chemical potential $\mu_{\cal O}$ as dictated by eq. \eqref{muqrelation}
\section{Scrambling time from CFT with higher spin charge} 
\label{scrambling_CFT}
In this section we turn to the CFT in the grand canonical ensemble with higher spin charge. Such a state is dual to a black hole with higher spin hair \cite{Gutperle:2011kf}. The backreacted solutions for infalling shockwaves which are generalizations of  \cite{Shenker:2013pqa} and \cite{Caputa1} to the charged higher spin case are not known. However, we can use existing  results from CFT to understand the time evolution of entanglement for certain situations where the CFT is held at finite higher spin chemical potential. In the Lagrangian formulation we view this as the (holomorphic) deformation\,,
\be
S_{\rm CFT}\,\to\,S_{\rm CFT}\,-\,\int d^2z\,\mu\,W(z)\,+\, {\rm h.c.}
\ee
where $\mu$ is the chemical potential for the higher spin current $W(z)$.

\subsection{Thermofield double and entanglement growth }
A particularly simple application involves the computation of the rate of growth of entanglement \cite{Calabrese:2005in} in the thermofield double state \cite{Hartman:2013qma}. We may consider a large interval of length $\ell \gg \beta$ (e.g. the half-line)  in the CFT and its thermofield double, and take time evolution to go forward in both copies. In a conformal field theory the entanglement entropy for the two copies $(L)$ and $(R)$ is expected to grow linearly at late times with an entanglement velocity $v=1$, and saturate at the thermal value at $t\simeq \ell/2$. The entanglement velocity is defined through the relation (for large intervals), 
\bea
S_{\rm EE}(L\cup R)\,\simeq\,4S_\beta\, v\, t\,.
 \eea
 where $S_\beta$ is the thermal entropy density. We first verify whether this relation is affected by the presence of a higher spin chemical potential.

To calculate the connected contribution to the entanglement entropy of the two copies of the interval, $L\cup R$, we need the  correlation functions twist field operators, each inserted at an endpoint of the two copies so that
\begin{equation}
S_{\rm EE}(L\cup R)\,=\,\sum_{i=1,2}\lim_{n\to 1}\,\frac{1}{n-1}\ln\left\langle {\cal T}_n (t_R, \ell_i)\, \widetilde{\cal T}_n (t_L,\,\ell_i)  \right\rangle^{{(\mu)}}\,.\label{leftright}
\end{equation}
The spatial endpoints of the two copies are at $\ell_1$ and $\ell_2$ with $\ell\,=\,\ell_2-\ell_1$.
The correlator is evaluated in the CFT perturbed 
by the chemical potential $\mu$ at finite temperature. In the spin-three case, this  was evaluated at $O(\mu^2)$ for equal times and unequal spatial points  in the thermal state \cite{universalcft}. The calculation generalises straightforwardly, by analytic continuation, to the situation when the twist operators are inserted 
at points $(z_1, \bar z_1)$ and $(z_2, \bar z_2)$. 
The result for the R\'{e}nyi entropy using this correlator  is given at order $\mu^2$ by,
\begin{eqnarray}  \label{renyi}
S^{(n)}\,=\,\frac{c (1+n)}{6n} &&\ln \left| \sinh\tfrac{\pi}{\beta}  (z_1 - z_2) \right|\,\\\nonumber
&&\hspace{1in} + \,\mu^2 \frac{ 5  c}{ 12 \pi^2 (n-1) } \left\{ 
{\cal S} (z_1 -z_2)   + {\cal S} (\bar z_1 - \bar z_2)  \right\} \,.
\end{eqnarray}
When $z_i$ and $\bar z_i$ are complex conjugates of each other, the expression is manifestly real. The function ${\cal S}$ is the result of integrating the four point correlator  $\sim\int d^2y_1\,d^2y_2 \langle {\cal T}_n\,W(y_1) W(y_2)\widetilde{\cal T}_n \rangle$ which arises at second order in the perturbation by the spin three current:
\begin{eqnarray}
{\cal S} ( z_1 -z_2)\, =\,  f_1\, {\cal I}_1 (z_1 - z_2)\,  +\, f_2 \,{\cal I}_2 (z_1-z_2)\,. \label{universalee}
\end{eqnarray}
with
\begin{eqnarray}
&&{\cal I}_1\left(z\right)\,=\, \frac{4\pi^4}{3\beta^2}\,\left(\frac{4\pi z}{\beta}\,\coth\left(\tfrac{\pi z}{\beta}\right)\,-\,1\right)\,+\,
\label{I1final}\\ \nonumber
&&\hspace{1.5in}+\,\frac{4\pi^4}{\beta^2}\sinh^{-2}\left(\tfrac{\pi z}{\beta}\right)\,\left\{\left(1-\frac{\pi z}{\beta}\coth\left(\tfrac{\pi z}{\beta}\right)\right)^2\,-\,\left(\tfrac{\pi z}{\beta}\right)^2\right\}\,,
\\\nonumber
&&{\cal I}_2 \left(z\right)\,=\,\frac{8\pi^4}{\beta^2}\,\left(5\,-\,\frac{4\pi z}{\beta}\,\coth\left(\tfrac{\pi z}{\beta}\right)\right)\,+\,
\label{I2final}\\  \nonumber
&&\hspace{1.5in}+\,\frac{72\pi^4}{\beta^2}\sinh^{-2}\left(\tfrac{\pi z}{\beta}\right)\,\left\{\left(1-\frac{\pi z}{\beta}\coth\left(\tfrac{\pi z}{\beta}\right)\right)^2\,-\,\frac{1}{9}\left(\tfrac{\pi z}{\beta}\right)^2\right\}\,,
\end{eqnarray}
and
\begin{equation}
f_1 \,=\, \frac{n^2-1}{4n}\,,\qquad \qquad f_2\, =\, \frac{(n^2 -1)^2}{120 n^3}\,-\,\frac{n^2 -1}{40 n^3}\,.
\end{equation}
The calculation of the connected left-right correlator \eqref{leftright} is then achieved by setting
%{\bf One sided correaltor}
%The prescription to find the equal space one sided correlator is 
%to take 
%\begin{equation} 
%(z_1, \bar z_1) = (   t_{R, 1 } , - t_{R, 1}) , \qquad 
%(z_2, \bar z_2)  = (   t_{R, 2 } , - t_{R, 2})
%\end{equation}
%and then substitute this into (\ref{renyi}). 
%{\bf Right-Left correlator}
%The prescription to find the right-left  equal space correaltor  is to take
\begin{equation}
(z_1^{(i)},\, \bar z_1^{(i)})\, =\, \left( \ell_i-t_{R}, \,\ell_i+ t_{R}\right)\, , \quad
\qquad(z_2^{(i)},\, \bar z_2^{(i)})\, =\, \left(\ell_i\,-\, t_{L}\, -\, i \tfrac{\beta}{2},  \,\ell_i \,+\,t_{L}\, +\, i \tfrac{\beta}{2}\right) \nonumber
\end{equation}
and then substituting into (\ref{renyi}). The resulting EE for the two copies of the half-line in the (forward) time evolved thermofield double state is,
%\begin{align}
%(z_1,\bar{z}_1) = (t_R,-t_R), \quad (z_2,\bar{z}_2) = (t_L+ %\frac{i \beta}{2},-t_L+ \frac{i \beta}{2})
%\end{align}
%in equation (8.2) is,
\bea
&&\tfrac{1}{2}S_{\rm EE}({L\cup R })\,=\,\frac{c}{3}  \ln\left[\cosh\tfrac{\pi}{\beta}  (t_L-t_R)\right]\\
&&-\,\frac{c\mu ^2 \pi ^2}{18 \beta ^4} {\rm sech}^4\tfrac{\pi}{\beta}  (t_L-t_R)\, \left[8 \pi \beta (t_L-t_R) \left(\sinh \tfrac{2 \pi}{\beta}  (t_L-t_R)\,-\,\sinh \tfrac{4 \pi}{\beta}  (t_L-t_R)\right)\right.  \nonumber\\
 && \left. +\,\left(8 \beta ^2\,+\,6 \pi ^2 \left(\beta ^2\,-\,4 (t_L-t_R)^2\right)\right) \cosh \tfrac{2 \pi}{\beta}  (t_L-t_R)\,+\,5 \beta^2  \cosh \tfrac{4 \pi}{\beta}  (t_L-t_R)\, +\, 3 \beta^2 \right].\nonumber
\eea
This result for the entanglement entropy can also be reproduced by the {\em holomorphic} Wilson line proposal \cite{universalcft} in the holographic dual, with one added subtlety. As usual  
for the ${\rm SL(3,{\mathbb R})}$ Wilson line we need to work in the adjoint representation where $W_{\rm Ad}\,=\,W_{\rm fund}\,W_{\overline{\rm fund}}$. For the two-sided Wilson line, the coordinates of the endpoints are complex. In order to obtain real results which match with the CFT prediction, the anti-fundamental Wilson lines also need to be complex conjugated.
Therefore, $W_{\rm fund}$ connects points $x_{R\pm}\,=\,t_R\pm x$ and $x_{L\pm}\,= \,t_L\,+\,\frac{i\beta}{2} \pm x$, while $W_{\overline{\rm fund}}$ is anchored to the points $x_{R\pm}\,=\,t_R\, \pm\, x$ and $x_{L\pm}^*\, =\, t_L\,-\,\frac{i\beta}{2} \pm x$.
It will be interesting to carry out this analysis to higher order in perturbation theory 
in the chemical potential 
using the results of \cite{Long:2014oxa}. 

\subsection{Scrambling time and entanglement speed} 
\paragraph{Small $\mu$:} Taking both copies of the thermofield double to have evolved forward for a time $t$,
we set $t_L-t_R\,=\,2t$ in our result for $S_{L\cup R}$ and then take the large $t$ limit:
\begin{align}
S_{\rm EE}({L \cup R})\to\,4\left(\frac{\pi c}{3\beta}\,+\,\frac{32 c\pi^3\mu^2}{9\beta^3}\right)\,t\,=\,4\,S_\beta\, t\,.
% \frac{4 t_* c \left(3 \pi  \beta ^2+32 \pi ^3 \mu ^2\right)}{9 \beta ^3} = 4 s t_*
\end{align}
where $S_\beta$ is precisely the thermal entropy density at order $\mu^2$ (see \cite{universalcft}).  This immediately implies that the entanglement speed $v=1$ at this order in $\mu$. In Einstein gravity this growth of entanglement is given a geometric interpretation in terms of the growth of the ``nice slice'' regions in the interior of the BTZ black hole  \cite{Hartman:2013qma}. In the presence of higher spin hair, it would be fascinating to understand how such notions are generalised, given that any geometric interpretation is  no longer obvious (see \cite{Castro:2016ehj} for investigations in this direction).

Let us note that the mutual information for the two large intervals is
\be
I_{L:R}\,\simeq\, 2 S_\beta\, \ell \,-\, S_{\rm EE}(L\cup R)\,.
\ee
On the right hand side we have replaced the entanglement entropy of each interval with the thermal entropy at large $\ell$. Therefore, the mutual information vanishes at $t_*\simeq \ell/2$, and remains zero beyond that point, since the disconnected contribution to $S_{\rm EE}(L\cup R)$ dominates.

\begin{figure}[h]
\begin{center}
\includegraphics[width=1.9in]{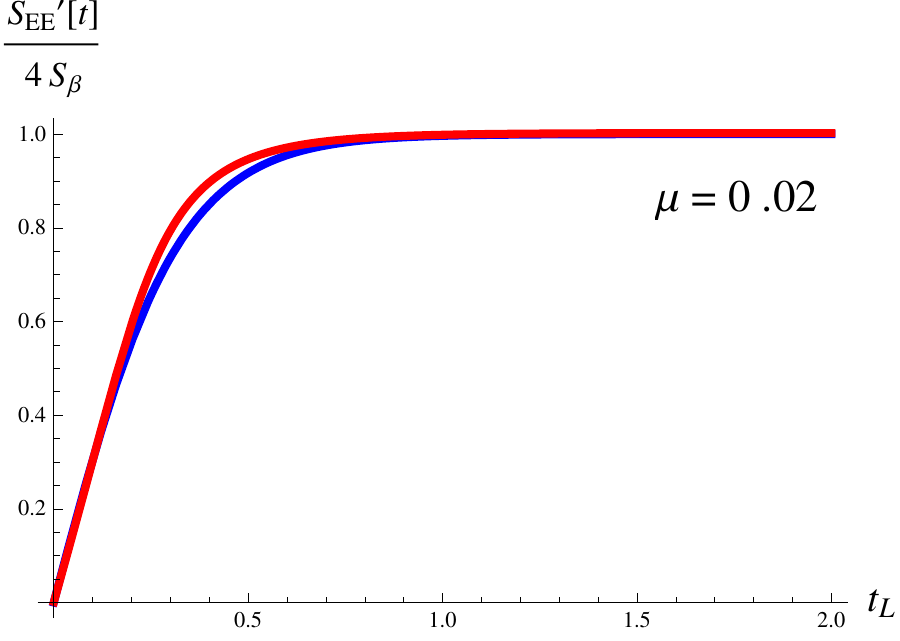}
\includegraphics[width=1.9in]{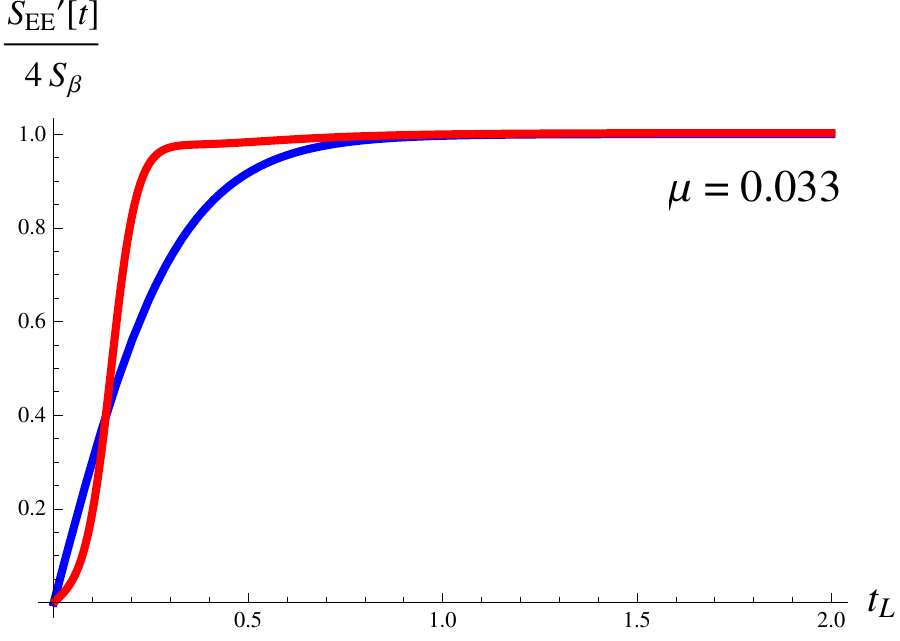}
\includegraphics[width=1.9in]{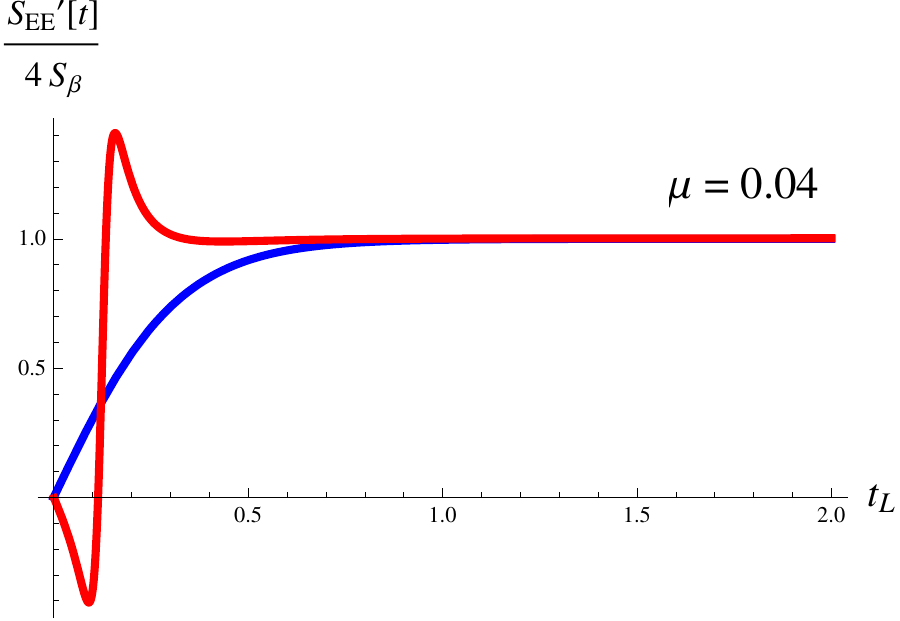}
\end{center}
\caption{\small{The  speed of entanglement entropy growth $v$ for three values of $\mu/\beta$ with $\beta =1$ for convenience, and taking the interval length $\ell \gg 1$. When $\mu > 0.033$, $v$ approaches unity from above. The blue curve is for $\mu = 0$ and the red curve is for $\mu \neq 0$.}}
\label{eegrowth}
\end{figure}

\begin{figure}[h]
\begin{center}
\includegraphics[width=2.3in]{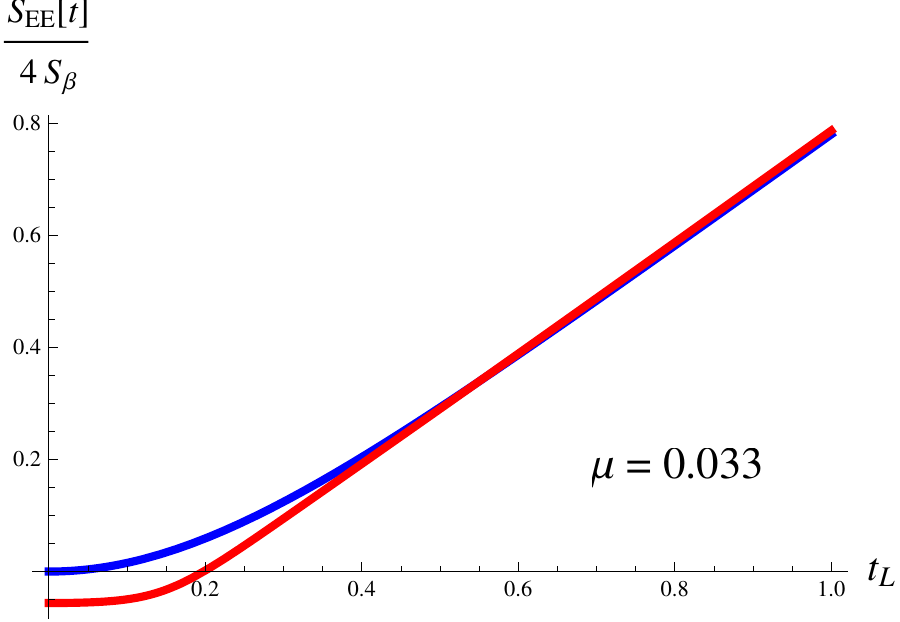}\hspace{1in}
\includegraphics[width=2.3in]{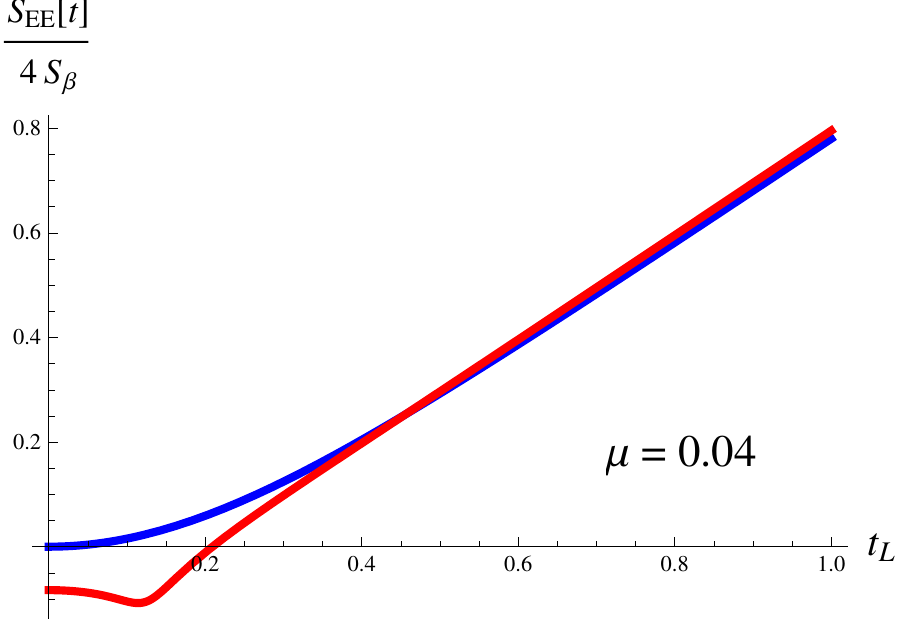}
\end{center}
\caption{\small{The entanglement entropy of $L\cup R$ as a function of time. The blue curve is for $\mu = 0$ and the red curve is for $\mu \neq 0$.}}
\label{eegrowth2}
\end{figure}

\paragraph{Exact result:} We know that the holomorphic Wilson line, appropriately defined, matches the CFT result for entanglement growth at order $\mu^2$, and also yields $v=1$ at late times. We can now use the  complete holomorphic Wilson line, to all orders in $\mu$, evaluated on the bulk spin three black hole solution, to calculate entanglement growth in the time evolved thermofield double state. The holomorphic component of the  Chern-Simons connection $a_+$ retains the form displayed in 
eq. \eqref{HSmu_connection}, and the expression for the holomorphic Wilson line in terms of the eigenvalues of $a_+$ is given by eq. \eqref{fullwad}. The values of the charges ${\cal L}$ and ${\cal W}$ are determined by the solutions to the black hole holonomy conditions and we pick the roots corresponding to BTZ-branch of \cite{David:2012iu}, which is smoothly connected to the BTZ black hole and dominates the grand canonical ensemble.
Crucially, the BTZ branch only exists for low values of $\mu/\beta$ and disappears above the critical value
\be
\frac{\mu_c}{\beta}\,=\,\frac{3}{16\pi}\sqrt{2\sqrt 3 -3}\,\approx\,0.04066\,.
\ee
In the absence of a clean closed form analytical expression, we plot the result for the entanglement growth numerically.  Figure \ref{eegrowth}  shows the rate of growth of  $S_{\rm EE}(L\cup R)$ as a function of time.
For low values of $\mu$ we see that $v$ approaches unity from below for large times, whereas when $\mu \gtrsim 0.033 \beta$, $dS_{\rm EE}/dt$ experiences an overshoot and eventually $v$ approaches $1$, but from above. Interestingly, the overshoot decreases and seems to disappear as $\mu$ approaches $\mu_c$. 
Thus there are times at which entanglement speed is greater than 
the speed of light which is the natural bound obtained 
in Einstein gravity in ${\rm AdS}_3$, see \cite{Erdmenger:2017gdk} for a review of these bounds. 
The significance of this is unclear. This
 behaviour for large times is  simultaneously 
accompanied by equally puzzling short time features. For $\mu \gtrsim 0.033 \beta $, the growth rate goes negative at early times, which means that 
$S_{\rm EE}({L\cup R})$ decreases first before growing linearly at late times and approaching the thermal value as shown in figure \ref{eegrowth2}.
Therefore $S_{\rm EE}({L\cup R})$ not a  monotonically increasing  function of 
time at early times 
 and large enough chemical potential\footnote{It is worth noting that the single-interval entanglement entropy given by the holomorphic Wilson line at $\mu\neq 0$ is a monotonically increasing function of the interval length,  which is consistent with the strong subadditivity property.}. This points to a violation of the strong 
 sub-additivity property of entanglement entropy  which has been observed 
 when the null energy condition is violated \cite{Allais:2011ys}. 
 Evaluating the mutual information, one sees that it is no longer a 
 concave function in time. 
 To conclude  there exists a bound on the chemical potential
 $\mu \sim 0.033 \beta $, below which 
 the mutual information and entanglement speed are well behaved.  

%\begin{align}
%s = \frac{\pi c}{3\beta} \left(1+\mu^2\frac{32 \pi^2 }{3\beta^2}\right).
%\end{align}
%Using the following definition of entanglement velocity,
%\begin{align}
%v = \frac{S_{A \cup B}}{4st_*}
%\end{align}
%we find the the entanglement velocity, $v=1$ in this setup.

\subsection{HHLL correlator at finite $\mu$}
While we do not have the  holographic description of a quench when the CFT is in the grand canonical ensemble with higher spin chemical potential, we can shed light on some aspects of such a setup from purely field theoretic considerations. In particular, we will argue that when the quenching operator ${\cal O}$ does {\em not} carry a higher spin charge, the change in the single interval entanglement entropy following the quench is unaffected by the higher spin chemical potential $\mu$, at order $\mu^2$. We will also find the change in the scrambling time due to the chemical potential at this order in $\mu$.

Let us consider the correlator between the two heavy $({\cal O}^\dagger, {\cal O})$ and the two light twist field operators which compute R\'enyi/entanglement entropies,
\begin{equation} \label{basiccol}
S_{n} (t) \,=\, \frac{ \langle {\cal O} ^\dagger(x_1, \bar x_1) \,{\cal T}_n( x_2 , \bar x_2) 
\widetilde{\cal T}_n( x_3, \bar x_3) \, {\cal O}(x_4, \bar x_4)  \rangle^{(\mu)}} 
{\left(\langle {\cal O}^\dagger ( x_1, \bar x_1) {\cal O} (x_4, \bar x_4) \rangle^{(\mu)}\right)^n }\,.
\end{equation}
The positions of the operator insertions are precisely those given in the beginning of this paper (see eq. \eqref{positions}). The correlators are evaluated on the thermal cylinder and the field theory is held 
at finite higher spin chemical potential $\mu$.  As usual, the  correlator above can be viewed as $2n$ insertions of the quenching operator  on the replica geometry. In the replica geometry the numerator of eq. (\ref{basiccol}) can be rewritten as,
\begin{equation}
\frac{ \langle {\cal O}^\dagger (x_1, \bar x_1) \,{\cal T}_n( x_2,  \bar x_2) 
\,\widetilde{\cal T}_n( x_3, \bar x_3)\, {\cal  O}(x_4, \bar x_4)  \rangle^{(\mu)} }{
\langle {\cal T}_n( x_2, \bar x_2) \,
\widetilde {\cal T}_n( x_3, \bar x_3)  \rangle^{(0)} }
\,=\, \langle 
\prod_{j=1}^n  {\cal O} ( x_4^{(j)}, \bar x_4^{(j)} )\, {\cal O} ^\dagger( x_1^{(j)} \bar x_1^{(j)} ) 
\rangle_n ^{(\mu)}\nonumber\,,
\end{equation}
where the denominator of the left hand side is evaluated at zero chemical potential. 
Now we may view  the right hand side as a perturbative expansion in  $\mu$, where each individual term is evaluated in the $\mu=0$ theory,
\begin{equation}
\langle 
\prod_{j=1}^n  {\cal O} ( x_4^{(j)}, \bar x_4^{(j)} )\, {\cal O} ^\dagger( x_1^{(j)} \bar x_1^{(j)} ) 
\rangle_n^{(\mu)} \, =\, 
{\cal C}^{(0)} \,+\, \mu \,{\cal C}^{(1)}\, +\, \frac{\mu^2}{2} \,{\cal C}^{(2)}\ldots
\end{equation}
${\cal C}^{(0)}$ is the $2n$-point correlator in the replica geometry for vanishing $\mu$, and the first putative correction ${\cal C}^{(1)}$ is given by 
\begin{equation}
{\cal C}^{(1)}\, =\, \sum_{i =1}^n \int d^2 y\, 
\langle  W(y^{(i)})\,  
\prod_{j=1}^n  {\cal O}( x_4^{(j)}, \bar x_4^{(j)} ) \,{\cal O}^\dagger ( x_1^{(j)} \bar x_1^{(j)}\rangle_n^{(0)}\,.
\end{equation}
If the operator ${\cal O}$ carries no charge under $W$, then the three-point functions essentially factorise on general grounds.  In particular, if ${\cal O}$ carries some charge ${\cal W}$ under $W$, then its OPE with $W$ takes the form
\begin{equation}
W(z)\, {\cal O}( x, \bar x)\, \sim\, \frac{{\cal W}}{ ( z - x)^3}{\cal O}(x,\bar x) \,.
\end{equation}
If we take ${\cal W}=0$, then  in the limit $\epsilon\rightarrow 0$,  the operators ${ \cal O}^\dagger$ and ${\cal O}$ are close and their OPE is dominated by the stress tensor  \cite{paper1}. Given that $W$ is a chiral primary, the correlator vanishes because the one-point function of $W$ in the $\mu=0$ theory must vanish. 
%The leading term is the disconnected correlator
%\begin{equation}
%\sum_{i =1}^n \int d^2 y 
%\langle  W(y^{(i)}  ) \rangle_n 
%\langle \prod_{j=1}^n  {\cal O}^\dagger ( x_4^{(j)}, \bar x_4^{(j)} ) {\cal O} ( x_1^{(j)} \bar x_1^{(j)}
%\rangle_n
%\end{equation}
%The $\epsilon^2$ term  in ${\cal C}^{(1)}$, which arises due to the presence of the 
%stress tensor  in the OPE of the primaries 
%also vanishes as analysed in the previous paper. 
The first non-vanishing contribution thus appears at order $\mu^2$, and in the 
small-$\epsilon$ limit, we find that the leading contribution again appears from the disconnected term:
%It is of the form 
%\begin{equation}
%{\cal C}^{(1)} \frac{1}{2} \sum_{i , k =1}^n \int d^2 y_1 d^2 y_2 
%\langle W(y_1^{(i) }) W (y_2^{(j) })
% \prod_{j=1}^n  {\cal O}^\dagger ( x_4^{(j)}, \bar x_4^{(j)} ) {\cal O} ( x_1^{(j)} %\bar x_1^{(j)}
%\rangle_n
%\end{equation}
%Now the $\epsilon^2$ dependence of this correlator was analysed in the previous paper. 
\begin{eqnarray}\label{c2}
{\cal C}^{(2)} &=& \frac{1}{2} \sum_{i , k =1}^n \int d^2 y_1 d^2 y_2 
\langle W(y_1^{(i) } )W (y_2^{(k)} ) \rangle_n  \times
\langle 
\prod_{j=1}^n  {\cal O} ( x_4^{(j)}, \bar x_4^{(j)} ) {\cal O}^\dagger ( x_1^{(j)} \bar x_1^{(j)}
\rangle_n^{(0)} \nonumber\\
& =& {\cal K} \times {\cal J}\,. 
\end{eqnarray}
The second factor, ${\cal J}$, is the $2n$-point correlator in the undeformed replicated theory, which can be rewritten in terms of the twist operator.
%We obtain 
%\begin{eqnarray}
%{\cal J} &=& \langle 
%\prod_{j=1}^n  {\cal O}^\dagger ( x_4^{(j)}, \bar x_4^{(j)} ) {\cal O} ( x_1^{(j)} \bar x_1^{(j)}
%\rangle_n  , \\ \nonumber
%&=& \frac{ \langle {\cal O} (x_1, \bar x_1) \sigma_n( x_2,  \bar x_2) 
%\tilde \sigma_n( x_3, \bar x_3) {\cal  O} ^\dagger(x_4, \bar x_4)  \rangle^{(0)} }{
%\langle \sigma_n( x_2 \bar x_2) 
%\tilde \sigma_n( x_3, \bar x_3)  \rangle^{(0)} }
%\end{eqnarray}
Since this is  evaluated at zero chemical potential, we may use the 
results of \cite{Caputa2} for the heavy-heavy-light-light limit 
of this correlator.  The first factor ${\cal K}$ in  eq.\eqref{c2} was computed exactly in \cite{universalcft} for a spin-three chemical potential.
Now we turn to the denominator in \eqref{basiccol}. Using similar arguments for small $\epsilon$ we get,
\begin{eqnarray}
& & \langle {\cal O}^\dagger ( x_1, \bar x_1) {\cal O} (x_4, \bar x_4) \rangle^{(\mu)} \\ \nonumber
& &=\,   \langle {\cal O}^\dagger ( x_1, \bar x_1) {\cal O}(x_4, \bar x_4) \rangle^{ (0)}
\left( 1 + \frac{\mu^2}{2}  \int d^2 y_1 d^2 y_2 
\langle W(y_1) W (y_2)\rangle\, +\ldots\right )\,.
\end{eqnarray}
Combining the $O(\mu^2)$ contributions of the numerator and denominator in 
eq.\eqref{basiccol} and taking the logarithms, and using the results of \cite{Caputa2} for the change in the undeformed CFT entanglement entropy, we find that  $\Delta S_{\rm EE}(t)$ following the local quench is unchanged from the CFT result. In particular, the factorisation seen above in the small-$\epsilon$ limit, due to the fact that the quench operator carries no higher spin charge, only changes the entanglement entropy of the interval by a constant, time independent amount:
\be
S_{\rm EE}(t)\,\simeq\,S_{\rm EE}^{(0)}(t)\,+\,\mu^2\,{\cal K} ( x_3, \bar x_3; x_2, \bar x_2)\,.
\ee
${\cal K}$ is independent of time, and for the case of the spin-three chemical potential, 
\be
{\cal K}   ( x_3, \bar x_3; x_2, \bar x_2)\,=\,  \lim_{n\rightarrow 1}
\frac{5   c}{12 \pi^2 (n-1) } \left\{
{\cal S}( x_3 - x_2)   \,+\, {\cal S} (\bar x_3 - \bar x_2) \right\}\,,
\ee
where the function ${\cal S}$ is defined in eq.\eqref{universalee}. As a result, at order $\mu^2$, the change in the single interval entanglement entropy following the local quench is, 
\begin{eqnarray} \label{caputa}
\Delta S_{\rm EE} && =\, 0\, \qquad\qquad t <\ell_1\,,\quad t >\ell_2 \,, 
\\\nonumber\\\nonumber
&& =\,   \frac{c}{6} 
\log \left[  \frac{\beta}{\pi \epsilon} \frac{\sin\pi \alpha\,
\sinh\tfrac{\pi}{\beta}(\ell_2 -t)\,\sinh\tfrac{\pi}{\beta}(t-\ell_1)}{\alpha \sinh\frac{\pi}{\beta}(\ell_2-\ell_1) }
 \right]\,,\qquad \ell_1 < t < \ell_2\,.
\end{eqnarray}
Here  $\alpha\,=\,\sqrt{1-24\Delta_{\cal O}/c}$. Therefore at quadratic order in $\mu$, a local quench by an operator ${\cal O}$ carrying no higher spin charge does not affect the time evolution of the single-interval entanglement entropy.

\subsection{Mutual information}

We may now apply the above analysis to the six-point  correlation function which computes the mutual information of two intervals $(L)$ and $(R)$, each in one copy of the CFT constituting the thermofield double state as in \cite{Caputa2}.  The correlator of interest is given by 
\begin{equation}
S^{(n)}  ({L\cup R}) 
\,= \,\frac{ \langle {\cal O }^\dagger(x_1, \bar x_1)\, {\cal T}_n(x_2, \bar x_2) \widetilde {\cal T}_n(x_3, \bar x_3) \,
{\cal T}_n(x_5, \bar x_5)\widetilde {\cal T}_n(x_6, \bar x_6) {\cal O} ( x_4, \bar x_4) \rangle^{(\mu)}  }
{ \langle {\cal O }^\dagger( x_1 \bar x_1) {\cal O} ( x_4, \bar x_4)  \rangle^{(\mu)} }\,.\nonumber
\end{equation}
The insertions points of two twist operators and the quenching operators on the left boundary are essentially the same as before, while the two remaining twist fields are inserted on the second boundary: 
\begin{eqnarray}
&&x_1\, =\, -i\epsilon, \qquad x_2\,=\,\ell_1 - t_L, \qquad x_3\, =\, \ell_2 - t_L\,, \qquad x_4\, =\, +i \epsilon\,, 
\\ \nonumber\\\nonumber
&& \bar x_1\, =\, +i\epsilon , \qquad \bar x_2\, =\, \ell_1 + t_L\,, \qquad 
\bar x_3 \,=\, \ell_2 + t_L\,, \qquad \bar x_4 = -i \epsilon, \\ \nonumber\\\nonumber
&& x_5 \,=\, \ell_2 + i \tfrac{\beta}{2} - t_R, \quad
x_6 = \ell_1+ i \tfrac{\beta}{2} - t_R,\quad \bar x_5 = \ell_2 - i \tfrac{\beta}{2} \,+\,t_R, \quad
\bar x_6 \,=\, \ell_1\,- \,i \tfrac{\beta}{2}\, +\, t_R\,.
\end{eqnarray}
All  correlators above are assumed to be evaluated at finite higher spin chemical potential $\mu$. We can now apply the S-channel and and T-channel factorizations in the large-$c$ picture which correspond to the disconnected and connected contributions from bulk gravity geodesics \cite{Caputa2}. From the arguments developed in the previous subsection, we conclude that at order $\mu^2$, the entanglement entropy of $L\cup R$ in the  S-channel is 
\begin{eqnarray}
S: \quad S^{(\mu)}_{\rm EE}({L\cup R}) &=& S^{(0)}_{\rm EE}(L)\,+\,S^{(0)}_{\rm EE}(R)
%\frac{ 2 c}{3} \log \left| \frac{\beta}{ \pi \epsilon_{UV} }
%\cosh( \frac{\pi \Delta t }{\beta} ) \right| + \mu^2 {\cal K} ( x_3 , x_2; \bar x_3, \bar x_2) 
%+\mu^2 {\cal K} ( x_6, x_5, \bar x_6, \bar x_5)  + t_- + t_\omega < y, 
%\\ \nonumber
%S_{A\cup B} &=& \frac{2c}{3} \log \left| \frac{\beta}{ \pi \epsilon_{UV} }
%\cosh( \frac{\pi \Delta t }{\beta} ) \right|     y< t_- + t_\omega <y+L \\ \nonumber
%&& + \frac{c}{6} \log \left( 
%\frac{\beta}{\pi\epsilon} 
%\frac{\sin\pi \alpha_\psi}{\alpha_\psi \cosh \frac{\pi \Delta t} {\beta} }
%\sinh \pi \frac{( t_- + t_\omega - y) }{ \beta} 
%\cosh \pi \frac{ ( t_+ + t_\omega -y)}{\beta} 
%\right) \\ \nonumber
 \,+\,  \mu^2 {\cal K} ( x_3 , \bar x_3; x_2, \bar x_2)
\,+\,\mu^2 {\cal K} ( x_6, \bar x_6; x_5, \bar x_5)\,.\nonumber
\end{eqnarray}
The order $\mu^2$ corrections are simply the contributions to the covariant single-interval EE at finite $\mu$.
Therefore the mutual information $I_{L:R}$ in this channel will continue to vanish at this order in the chemical potential. In the T-channel or ``connected" picture on the other hand one obtains,
\be
T:\quad S^{(\mu)}_{\rm EE}({L\cup R}) = S^{(0)}_{\rm EE}({L\cup R})\,+\, \mu^2 {\cal K} ( x_3 , \bar x_3; x_5, \bar x_5)
\,+\,\mu^2 {\cal K} ( x_6, \bar x_6; x_2, \bar x_2)\,.
\ee
The mutual information in this channel is not only non-zero but in fact the order $\mu^2$ correction is also non-vanishing,
\bea
&&I^{(\mu)}_{L:R}\,=\,I^{(0)}_{L:R}\,-\,\mu^2\left[{\cal K} ( x_3 , \bar x_3; x_5, \bar x_5)
\,+\, {\cal K} ( x_6, \bar x_6; x_2, \bar x_2)\right.\\\nonumber
&&\left.\hspace{2.5in}-\,{\cal K} ( x_3 , \bar x_3; x_2, \bar x_2)
\,-\,{\cal K} ( x_6, \bar x_6; x_5, \bar x_5)\right]\,.
\eea
%Now lets look at the T-channel. We have 
%\begin{eqnarray}
%S_{A\cup B} &=& \frac{c}{3} \log \left( 
%\frac{\sinh \frac{\pi ( t_- + t_\omega -y) }{\beta} \cosh \frac{\pi( t_+  +t_\omega -y) }{\beta} }{
%\cosh \frac{\pi ( \Delta t)}{\beta} }
%\frac{\sinh \frac{\pi ( t_- + t_\omega -y -L) }{\beta} \cosh \frac{\pi( t_+  +t_\omega -y-L) }{\beta} }{
%\cosh \frac{\pi ( \Delta t)}{\beta} }
%\right) \nonumber \\
%&& + \frac{2c}{3} \log \left| \frac{\beta}{ \pi \epsilon_{UV}} \cosh( \frac{\pi \Delta t}{\beta} )
%\right| \\ \nonumber
%&& + \frac{c}{3} \log \left( \frac{\beta}{\pi \epsilon} \frac{\sin \pi \alpha_{\psi}}{\alpha_\psi} \right)
%\\ \nonumber
%&& + \mu^2{\cal K} ( x_5, x_3; \bar x_5, \bar x_3) + \mu^2{\cal K} ( x_6, x_2; \bar x_6, \bar x_2) 
%\\ \nonumber
%& & t_- + t_\omega > y+L
%\end{eqnarray}
%Here the $\mu^2$ correction does not cancel with the $\mu^2$ correction of 
%$S_A + S_B$ which  depends on $L$. 
%Using this expression we can evaluate the correction to scrambling time from CFT. 
\subsection{Scrambling time}
It is now a straightforward exercise to calculate the scrambling time and pinpoint the corrections that arise from the higher spin chemical potential for the spin-three case where explicit results are available.  To calculate the scrambling time we first set $t_L=t_R=t$. With this choice we note that the order $\mu^2$ corrections to the mutual information become time independent:
\be
I^{(\mu)}_{L:R}(t)\,=\,I^{(0)}_{L:R}\,+\,\mu^2\,g(\ell,\,\beta)\,,
\qquad \ell\,\equiv\,\ell_2-\ell_1\,,\qquad t_L=t_R=t\,,
\ee
where, using the definition \eqref{universalee} of ${\cal K}$,
\bea
&&g(\ell,\,\beta)\,=\,2\,S^{(2)}_{\rm EE}\,+\,\frac{2c\pi^2}{9\beta^2}\,(8\,+\,3\pi^2)\,,\\\nonumber\\\nonumber
&& S^{(2)}_{\rm EE}\,=\,\frac{2c}{\beta^2}\left[\tfrac{32\pi^2}{9}\,\tfrac{\pi \ell}{\beta}\,{\rm coth}\tfrac{\pi \ell}{\beta}\,-\,\tfrac{20\pi^2}{9}\,-\,\frac{4\pi^2}{3\sinh^2\frac{\pi \ell}{\beta}}\left\{\left(\tfrac{\pi \ell}{\beta}\,{\rm coth}\tfrac{\pi \ell}{\beta}-1\right)^2\,+\,\left(\tfrac{\pi \ell}{\beta}\right)^2\right\}\right]\,.
\eea
Taking the large $t$ limit, we find that the mutual information vanishes at $t=t_*$,
\be
t_*\,=\,\frac{\ell_2+\ell_1}{2}\,+\,\frac{\beta}{2\pi} \frac{6 S_{\rm EE}(\ell)}{c}\,+\,\frac{\beta}{2\pi}\left(\ln\left(\frac{S_\beta^{(0)}}{\pi E}\right)\,+\,\frac{16\mu^2\pi^2}{3\beta^2}\,+\,2\pi^4\frac{\mu^2}{\beta^2}\right)
\ee
The single interval entanglement entropy, $6S_{\rm EE}(\ell)/c\,=\,\ln \sinh^2\frac{\pi \ell}{\beta}\,+\,\mu^2 S_{\rm EE}^{(2)}$, includes the first nontrivial correction at order $\mu^2$ while,
$S_\beta^{(0)}\,=\,\pi c/3\beta$ is the conformal result for thermal entropy density. The correction to the thermal entropy density at order $\mu^2$  is known to be \cite{universalcft},
\be
S_\beta\,=\,\frac{\pi c}{3\beta}\left(1\,+\,\frac{32\mu^2\pi^2}{3\beta^2}\,+\,\ldots\right)\,.
\ee 
However, the extra terms in the scrambling time are not completely accounted for by this shift in thermal entropy.  It would be interesting to understand the physical interpretation of the corrections. The coefficient of the term $\sim \ln S_\beta$ can be identified with the Lyapunov exponent. We conclude that for perturbations carrying no spin-three charge, the Lyapunov exponent in the spin-three black hole background is the same as that for pure gravity.

\section{Discussions}
The main results of this work are the identification of certain features in physical observables computed in a particular higher spin theory which extends gravity by the inclusion of one spin three field. One of these features appears in the single interval entanglement entropy following a local quench by an operator with spin three charge. For large enough charge the entanglement entropy  becomes unbounded from below (turning complex beyond that value). A similar feature appears in the time dependence of the entanglement entropy in the forward evolved thermofield double state when the CFT is held at finite chemical potential for spin three charge. However, in both cases it appears that for sufficiently small charge or chemical potential, the physical observables in question are well behaved. This is also reflected in the scrambling time for mutual information following the local quench. It is only beyond a critical value of the spin three charge, that the 
scrambling time decreases and the Lyapunov index assumes the spin 3 value. Therefore, if we want the chaos bound to be respected, and avoid accompanying pathological features,
there must be bounds on the higher spin charges carried by states in the putative CFT dual. As remarked earlier, our bound \eqref{critical} is consistent with the bounds obtained on the general grounds of unitarity and modularity for $W_N$ theories with $N$ fixed in \cite{Afkhami-Jeddi:2017idc}.

In this work we have focussed on the CFT with a single higher spin current described 
holographically by the ${\rm SL}(3,{\mathbb R})\times {\rm SL}(3,{\mathbb R})$
Chern-Simons theory. It will be interesting to  see how  the observations 
in this paper generalise 
 to the  ${\rm SL}(N,{\mathbb R})\times {\rm SL}(N,{\mathbb R})$ theory
 and  obtain the $N$ dependence of 
  the bounds we have found on the higher spin charge and the 
 chemical potential. 
  From the analysis of \cite{Perlmutter:2016pkf} using  CFT's with ${\cal W}_\infty$ symmetry, 
  it is expected that the 3d  Vasiliev theory with ${\rm hs}[\lambda]$
 does not exhibit chaos. It will be interesting to see this directly 
 in holography.  
 
 One situation we did not consider in this paper, is  when the CFT is held at 
 finite higher spin chemical potential and the local quench is induced 
 by an operator which also has higher spin charge. 
  It will be interesting to generalise the observations of this paper to this situation. 
  This will  involve the study of ${\cal W}_3$ conformal blocks deformed  
  by higher spin chemical potential possibly using the methods of \cite{Chen:2016uvu}. 
  
  Finally we mention that there is a   conceptually  simple picture to 
  obtain the back reacted geometry of the infalling particle and then 
  study the Lyapunov exponent and the scrambling time in 
  Einstein gravity \cite{Shenker:2013pqa}. This involves two BTZ geometries of different 
  masses sewed along a null shell representing the infalling particle and 
  then evaluating geodesics in this geometry. 
  We have reviewed  this approach using  Wilson lines instead of geodesics  
   in appendix \ref{app:infallingCS}.  
   It will be useful to generalise this picture when the 
   null shell also carries higher spin charge.  This will involve coupling 
   particles to higher spin Chern-Simons theories and obtaining 
   appropriate junction conditions across the null shell
   \footnote{We thank Julian Sonner for discussions regarding these
   issues.}. 
   The behaviour of scrambling time and Lyapunov index 
    in this picture is expected to coincide with the results obtained in this paper. 
    However performing this calculation in the conceptually simple picture
    will be useful to obtain a gauge invariant formulation of the junction conditions 
    and the higher spin shock wave 
    in the Chern-Simons language.

\acknowledgments  
We would like to thank Shouvik Datta, Daniel Grumiller  and Julian Sonner for discussions. We would also like to thank the Galileo Galilei Institute for Theoretical Physics (GGI) for hospitality and providing a stimulating atmosphere, and INFN for partial support during the completion of this work, within the program ``New Developments in ${\rm AdS}_3/{\rm CFT}_2$ Holography".  SPK acknowledges financial support from STFC grant ST/L000369/1.

\appendix
%\begin{appendix}
\section{Transforming conical deficit to infalling shell}
\label{app:fullshockwave}
In \cite{Caputa1} the transformations which map the conical deficit \eqref{staticdeficit} in ${\rm AdS}_3$ to an infalling particle in the BTZ background were obtained. The transformations map the centre of the  conical deficit to an infalling geodesic in the BTZ background \cite{Nozaki:2013wia}. The mass of the infalling particle in the bulk is taken to be 
$m\,=\, \Delta_{\cal O}/R $ where $\Delta_{\cal O}$ is the dimension of the operator producing the quench. 
We quote the transformations for completeness:
\begin{align}\label{fullshock}
\sqrt{r^2+R^2}\,\sin \tau &\, =\, \frac{R}{\sqrt{M}z}\,\sqrt{1-Mz^2}\,\sinh\left(\sqrt{M}t\right)\,,\\\nonumber\\
\sqrt{r^2+R^2}\,\cos {\tau} & \,=\, \frac{R}{\sqrt{M}z}{ \left[\cosh(\lambda)\cosh\left(\sqrt{M}x\right)-\sqrt{1-Mz^2}\sinh(\lambda)\cosh\left(\sqrt{M}t\right) \right]}\,,\nonumber\\\nonumber\\
r \sin(\phi) & \,=\, \frac{R}{\sqrt{M}z}\sinh\left(\sqrt{M}x\right)\,,\nonumber\\\nonumber\\
r\cos(\phi)& \,=\, \frac{R}{\sqrt M z}{ \left[\cosh(\lambda)\sqrt{1-Mz^2}\,\cosh\left(\sqrt{M}t\right)\,-\,\sinh(\lambda)\cosh\left(\sqrt{M}x\right) \right]}\,.\nonumber
\end{align}
Here $(t,x)$  are boundary CFT coordinates and $z$, the radial coordinate (AdS boundary at $z=0$) in the ordinary BTZ black hole geometry which is obtained from the above transformations on global AdS$_3$ without a defect ($m\,=\,0$):
\bea
ds^2\, =\, \frac{R^2}{z^2}\, \left(-(1-M z^2)\,dt^2\,+\,\frac{dz^2}{(1-Mz^2)}\,+\,dx^2 \right)\,.
\eea
The BTZ black hole mass is given by $M$ which is in turn related to the Hawking temperature as $\beta\,=\, 2\pi/\sqrt{M}$.
The parameter $\lambda$ generates a boost  which is a symmetry of the background in the absence of the defect. When $ m\neq 0$, the transformed geometry is time dependent and the boost parameter is linked to the width ($\tilde \epsilon$) of the excitation via
\begin{align}
\tanh \lambda \,=\, \sqrt{1-M\tilde\epsilon^2}\,.
\end{align}

\section{The Regge limit for chaos}
\label{app:zrot}
\begin{figure}[h]
\begin{center}
\includegraphics[width=2in]{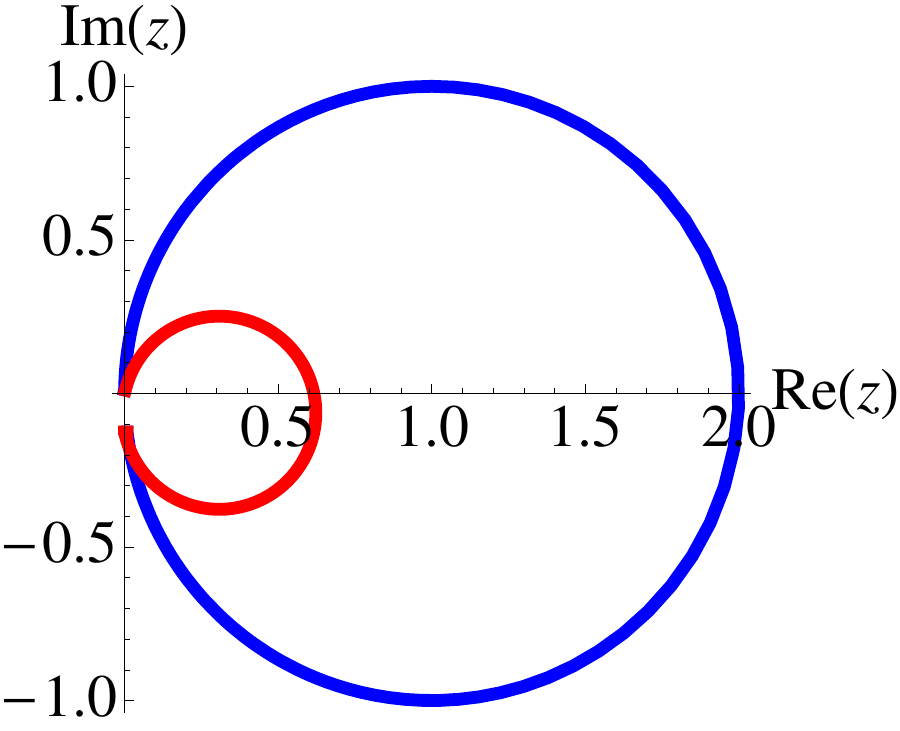}
\end{center}
\caption{\small{The path (in blue) followed by $z(t)$ as $t$ is dialled from $0$ to $(\ell_1+\ell_2)/2$ with a quench width $\epsilon=0.005$, $\beta=0.5$, $\ell_1=0.2$ and $\ell_2=1.8$. The continuation of the same for $t> (\ell_1+\ell_2)/2$ is performed by 
introducing an imaginary part for $t$, so that $t\to t+i\epsilon_1$ where  $ \epsilon_1 \approx 0.02$. The resulting path is shown in red. Importantly, the latter path does not encircle $z=1$, and approaches $z=0$ for large $t$.}}
\label{fig:oto2chaos}
\end{figure}

Let us consider the cross-ratio $z$, as defined in \eqref{zdef}. As $t$ increases from $0$, past $t=\ell_1$, to the deep interior of the interval $PQ$, the cross ratio traverses along the circle $|z-1|=1$ moving to the second sheet 
(figure \ref{fig:sl2ee}) of the Wilson line correlator viewed as a function of $z$. In order to access the chaos regime of large $t$ whilst staying on the second sheet, at time $t=(\ell_1+\ell_2)/2$ we introduce an imaginary part for  $t$, so that $t\to t+ i\tilde \epsilon(t)$. The imaginary component $\tilde \epsilon(t)$ is switched on smoothly at $t=(\ell_1+\ell_2)/2$ and increased to a constant asymptotic value larger than $\epsilon$, the width of the local quench. The result of this process is shown in figure \ref{fig:oto2chaos}. 
If $\epsilon_1$ is not introduced, for time $t>(\ell_1+\ell_2)/2$, the cross-ratio $z$ will retrace its path along the circle $|z-1|=1$ and return to the first sheet, at which point the excitation would exit the interval.
\section{Generators of ${\rm SL}(3, {\mathbb R})$}
\label{app:sl3}
The ${\rm sl}(3,{\mathbb R})$ algebra is generated by $L_0, L_{\pm 1}, W_0, W_{\pm1},
W_{\pm 2}$. Of these, $\{L_0, L_{\pm 1}\}$ generate the ${\rm sl}(2,{\mathbb R})$ 
subalgebra corresponding to pure gravity:
\be
[L_{i}, L_j]\,=\, (i-j)L_{i+j}\,, \qquad i,j\,=\,0,\pm 1\,.
\ee
The remaining commutation relations are,
\bea
&&[L_i, W_m]\,=\,(2i-m) W_{i+m}\,,\qquad m,n = 0, \pm1, \pm2\,,\\\nonumber
&&[W_m, W_n]\,=\,-\frac{1}{3}(m-n)(2m^2 + 2n^2-mn -8) L_{m+n}\,.
\eea
\section{Adjoint Wilson line in ${\rm SL(3,{\mathbb R})\times SL(3,\mathbb R)}$  C-S theory}
\label{app:wl}
The Wilson line in the adjoint representation is obtained form the product of the fundamental and anti-fundamental Wilson lines:
\be
W_{\rm Ad}\,=\,W_{\rm fund}\,W_{\rm \overline{fund}}\,.
\ee
To obtain the entanglement entropy we need to evaluate $W_{\rm fund}$ in the limit $\rho_{P,Q} \to \infty$ and retain the leading term, taking the barred and unbarred eigenvalues to be different for generality:
\bea
&&\lim_{\rho\to\infty}W_{\rm fund}\,=\,\lim_{\rho\to\infty}{{\rm Tr} \left[e^{L_0\rho_Q}\,e^{- \bar a_-\Delta \xi^-}\,e^{-2 L_0\rho_P}\,e^{ a_+\Delta \xi^+}\,e^{L_0\rho_Q}\right]}\\\nonumber\\\nonumber
&&=\, \frac{e^{2(\rho_P+\rho_Q)}}{\lambda_1\lambda_2\lambda_3\bar\lambda_1\bar\lambda_2\bar\lambda_3}\,\left(\lambda_1\,e^{\nu_3 \Delta \xi^+}\,+\,
\lambda_2\,e^{\nu_1 \Delta \xi^+}\,+\,\lambda_3\,e^{\nu_2 \Delta \xi^+}\right)\times\\\nonumber
&&\left(\bar\lambda_1\,e^{-\bar\nu_3 \Delta \xi^-}\,+\,
\bar\lambda_2\,e^{-\bar\nu_1 \Delta \xi^-}\,+\,\bar\lambda_3\,e^{-\bar\nu_2 \Delta \xi^-}\right)\,.
\eea
The Wilson line in the conjugate representation is given by the same expression involving  the barred connections, but with the signs of $\Delta \xi^\pm$ reversed:
\bea
&&\lim_{\rho\to\infty}W_{\overline{\rm fund}}\,=\,\lim_{\rho\to\infty}{{\rm Tr} \left[e^{L_0\rho_Q}\,e^{\bar a_-\Delta \xi^-}\,e^{-2 L_0\rho_P}\,e^{- a_+\Delta \xi^+}\,e^{L_0\rho_Q}\right]}\\\nonumber\\\nonumber
&&=\, \frac{e^{2(\rho_P+ \rho_Q)}}{\lambda_1\lambda_2\lambda_3\bar\lambda_1\bar\lambda_2\bar\lambda_3}\,\left(\lambda_1\,e^{-\nu_3 \Delta \xi^+}\,+\,
\lambda_2\,e^{-\nu_1 \Delta \xi^+}\,+\,\lambda_3\,e^{-\nu_2 \Delta \xi^+}\right)\,\times\\\nonumber\\\nonumber
&&\left(\bar\lambda_1\,e^{\bar\nu_3 \Delta \xi^-}\,+\,
\bar\lambda_2\,e^{\bar\nu_1 \Delta \xi^-}\,+\,\bar\lambda_3\,e^{\bar\nu_2 \Delta \xi^-}\right)\,.
\eea
Multiplying out the expressions for $W_{\rm fund}$ and $W_{\overline{\rm fund}}$, we obtain  
\bea
\lim_{\rho_{P, Q}\to\infty} &&W_{\rm Ad}(P,Q)\,=\,\frac{64\,e^{4(\rho_P + \rho_Q)}}{\left(\lambda_1\lambda_2\lambda_3\right)^2}\times\label{fullwad}\\\nonumber\\\nonumber
&&
\left[\frac{\lambda_1^2+\lambda_2^2+\lambda_3^2}{2\lambda_1\lambda_2\lambda_3}\,+\,\frac{\cosh\left(\lambda_1\Delta \xi^+\right)}{\lambda_1}\,+\,\frac{\cosh\left(\lambda_2\Delta \xi^+\right)}{\lambda_2}\,+\,\frac{\cosh\left(\lambda_3\Delta \xi^+\right)}{\lambda_3}\right]\times
\\\nonumber\\\nonumber
&&\left[\frac{\lambda_1^2+\lambda_2^2+\lambda_3^2}{2\lambda_1\lambda_2\lambda_3}\,+\,\frac{\cosh\left(\lambda_1\Delta \xi^-\right)}{\lambda_1}\,+\,\frac{\cosh\left(\lambda_2\Delta \xi^-\right)}{\lambda_2}\,+\,\frac{\cosh\left(\lambda_3\Delta \xi^-\right)}{\lambda_3}\right]\,.
\eea
Using $\lambda_1+\lambda_2+\lambda_3=0$ yields the form displayed in eq. \eqref{WLadj}.

\section{Gauge transformation of the BTZ connection} \label{appendix:gtbtz}
In this appendix, we explore the coordinate transformation from the conical deficit in global AdS to BTZ viewed as a gauge transformation in Chern-Simons theory. First we consider the coordinate transformation \eqref{fullshock} in the zero boost limit $(\lambda = 0)$ (see also eq.\eqref{AdS3toBTZ}),  and apply it to the global AdS connection, $A^{\rm AdS}(x)$,
\begin{align}
A'_{\mu}(x') \,=\, \frac{\partial x^\nu }{\partial {x'}^\mu} A^{\rm AdS}_\nu (x(x'))\,.
\end{align}
The connection $A'(x')$ is then a BTZ connection. However, it is a gauge transformed version of the usual BTZ connection $A^{(0) \rm BTZ}(x')$ (equation \eqref{usualBTZconnec}). We want to evaluate the gauge transformation $U$ which relates these two connections,
\begin{align} \label{gauge_trans0}
A'(x')\, =\, U A^{(0)\rm BTZ}(x') U^{-1}\, -\, dU \, U^{-1}\,.
\end{align}
Below we will be able to evaluate the gauge transformation at ${O}\left( \frac{1}{\hat{r}^2}\right)$ in the large-$\hat r$ expansion. In the following subsection, we will apply the same coordinate transformation \eqref{AdS3toBTZ} to the conical defect connection,
\begin{align}
\tilde{A}_{\mu}(x')\, =\, \frac{\partial x^\nu }{\partial {x'}^\mu} A^{\rm con}_\nu (x(x'))\,.
\end{align}
This gives a BTZ connection, $\tilde{A}(x')$. We then assume that the same gauge transformation in eq.\eqref{gauge_trans0}, relates $\tilde{A}(x')$ to the following BTZ connection,
\begin{align} 
\,\tilde{A}(x')\, =\, U A^{\rm BTZ}(x') U^{-1}\, -\, dU \, U^{-1},
\end{align}
where
\begin{align}
A^{\rm BTZ} \,=\, A^{(0) \rm BTZ} \,+\, \Delta A^{\rm BTZ}\,.
\end{align}
The above equations are used to determine $\Delta A^{\rm BTZ}$, which is given in equation \eqref{BTZ_A_corr}. This essentially gives the conical defect correction $\sim {O}(\delta)$ to the usual BTZ connection. In the final subsection of this appendix, we repeat the above steps for the case when the coordinate transformation has non-zero boost, $(\lambda \neq 0)$ (equation \eqref{coord_trans}) which corresponds to a finite width quench. Again, we derive the gauge transformation $U$ at order ${O}\left( \frac{1}{\hat{r}^2}\right)$. 
%However, the analysis is not carried out further to evaluate $\Delta A^{BTZ}$.

\subsection{AdS$_3$ to BTZ}
Consider the transformation from the AdS$_3$ metric,
\begin{align}
ds^2 = -(r^2+R^2)d\tau^2 + \frac{R^2 dr^2}{r^2+R^2}+r^2 d\phi^2,
\end{align}
to the BTZ metric,
\begin{align}
ds^2 = R^2 \left( - \left(\hat{r}^2 - M \right)dt^2 + \frac{d\hat{r}^2}{\hat{r}^2-M}+\hat{r}^2 dx^2\right),
\end{align}
where, {\bf $\hat{r}$ has dimensions of inverse length.}
The following transformation relates the AdS$_3$ coordinates to the BTZ coordinates when boost is zero,
\bea \label{AdS3toBTZ}
&&r = \frac{R \sqrt{\left(\hat{r}^2-M\right) \cosh ^2\sqrt{M} t+\hat{r}^2 \sinh ^2\sqrt{M} x}}{\sqrt{M}},\\\nonumber
&&\tau  = \tan ^{-1}\left(\frac{\sqrt{\hat{r}^2-M} \sinh \sqrt{M} t}{\hat{r} \cosh \sqrt{M} x}\right)\,,\qquad
\phi  = \tan ^{-1}\left(\frac{\hat{r} \sinh \sqrt{M} x}{\sqrt{\hat{r}^2-M} \cosh \sqrt{M} t}\right). \nonumber 
\eea
The AdS$_3$ connections,
\bea
&&A^{\rm AdS}_r  =\frac{L_0}{\sqrt{r^2+R^2}}\,,\qquad
A^{\rm AdS}_\tau  =\frac{L_{-1} R^2 \exp (-\rho ) + L_{1} \exp (\rho )}{2 R}\,,\\ 
&& A^{\rm AdS}_\phi  = \frac{L_{-1} R^2 \exp (-\rho ) + L_{1} \exp (\rho )}{2 R}\,, \nonumber 
\eea
are transformed to the BTZ connection using the transformation in equation \eqref{AdS3toBTZ},
\bea
A'_{\mu}(x') = \frac{\partial x^\nu }{\partial {x'}^\mu} A^{\rm AdS}_\nu (x(x'))\,,
\eea
where $x=(\tau,r,\phi)$ and $x'=(t,\hat{r},x)$. Here, the connections $A'_\mu(x')$ are related to the usual BTZ connections $A^{(0)\rm BTZ}_\mu(x')$\,,
\bea\label{usualBTZconnec}
A^{(0)\rm BTZ}_{\hat{r}}  \,=\, \frac{1}{\sqrt{\hat{r}^2 - M}} L_0\,,\quad
A^{(0)\rm BTZ}_t \,=\, A^{(0)\rm BTZ}_\phi =  \frac{1}{2} \left(e^{\hat{\rho}} L_1 - M e^{-\hat{\rho}} L_{-1} \right)\,,
\eea
by a gauge transformation, $U$, such that 
\begin{align} \label{gauge_trans}
A'(x')\, =\, U A^{(0)\rm BTZ}(x') U^{-1} \,-\, dU \, U^{-1}\,.
\end{align}
The matrix $U$ has determinant $1$. The following gauge transformation is obtained as a series expansion in $\frac{1}{\hat{r}}$,
\begin{align} \label{gauge_trans_matrix}
U = \begin{pmatrix}
u^{(0)}_{11}+\frac{u^{(2)}_{11}}{\hat{r^2}} & \frac{u^{(1)}_{12}}{\hat{r}}\\
\frac{u^{(1)}_{21}}{\hat{r}} & u^{(0)}_{22}+\frac{u^{(2)}_{22}}{\hat{r^2}}
\end{pmatrix} + \mathrm{O}\left(\frac{1}{\hat{r}^3}\right),
\end{align}
where,
\begin{align}
u_{11}^{(0)} & = 2^{-\frac{1}{4}} \sqrt{\text{sech}\sqrt{M}(t-x)}, \quad u_{22}^{(0)} = \frac{1}{u_{11}^{(0)}}\,,\\
u_{12}^{(1)} & = \frac{\sqrt{M}}{2}\frac{\sinh\sqrt{M}(t+x)}{\cosh^{\frac{3}{4}}\sqrt{M}(t+x)\cosh^{\frac{1}{4}}\sqrt{M}(t-x)}\,, \nonumber\\
u_{21}^{(1)} & = \frac{\sqrt{M}}{2}\frac{\sinh\sqrt{M}(t-x)}{\cosh^{\frac{1}{4}}\sqrt{M}(t+x)\cosh^{\frac{3}{4}}\sqrt{M}(t-x)}\,, \nonumber\\
u_{11}^{(2)} & = \frac{M}{2^5} \frac{\sinh\sqrt{M}(t+x)\left[3\sinh\sqrt{M}(t-3x)-\sinh\sqrt{M}(3t+x)\right]}{\cosh^{\frac{7}{4}}\sqrt{M}(t+x)\cosh^{\frac{9}{4}}\sqrt{M}(t-x)}\,, \nonumber\\
u_{22}^{(2)} & = \frac{M}{2^3} \frac{\sinh\sqrt{M}(t+x)\sinh\sqrt{M}(t-x)}{\cosh^{\frac{5}{4}}\sqrt{M}(t+x)\cosh^{\frac{3}{4}}\sqrt{M}(t-x)}\,. \nonumber
\end{align}
The above gauge transformation was obtained by demanding that the dreibein,
\begin{align}
e'_\mu = \frac{1}{2}(A'_\mu - \bar{A}'_\mu)\,,
\end{align}
transforms homogeneously under the gauge transformation,
\begin{align}
e'(x') = U e(x') U^{-1}\,,
\end{align}
where $e(x')$ is the dreibein corresponding to the usual BTZ connection of equation \eqref{usualBTZconnec}. The dreibeins as series expansion around $\hat{r} \to \infty$ are,
\begin{align}\label{veir_BTZ_prime}
e'_t & = \begin{pmatrix}
a_t & -b \hat{r} + \frac{c(x)}{\hat{r}}\\
\frac{b^{-1}}{4} \hat{r}-\frac{c(-x)}{\hat{r}} & -a_t
\end{pmatrix}, \quad  
e'_x  = \begin{pmatrix}
a_x & b \hat{r} - \frac{a_x d(x)}{\hat{r}}\\
\frac{b^{-1}}{4} \hat{r}+\frac{a_x d(-x)}{\hat{r}} & -a_x
\end{pmatrix},\\
e'_{\hat{r}}&=\left(
\begin{array}{cc}
 \frac{1}{2 \hat{r}} & - \frac{d(x)}{\hat{r}^2}\\
\frac{d(-x)}{\hat{r}^2} & -\frac{1}{2 \hat{r}} \\
\end{array}
\right),\nonumber
\end{align}
and,
\begin{align}
e_{\hat{r}} & = \left(
\begin{array}{cc}
 \frac{1}{2 \sqrt{\hat{r}^2-M}} & 0 \\
 0 & -\frac{1}{2 \sqrt{\hat{r}^2-M}} \\
\end{array}
\right) \xrightarrow{\hat{r} \to \infty} \left(
\begin{array}{cc}
 \frac{1}{2 \hat{r}} & 0 \\
 0 & -\frac{1}{2 \hat{r}} \\
\end{array}
\right),\\
e_t & = \left(
\begin{array}{cc}
 0 & -\frac{1}{2} \sqrt{\hat{r}^2-M} \\
 \frac{\sqrt{\hat{r}^2-M}}{2} & 0 \\
\end{array}
\right) \xrightarrow{\hat{r} \to \infty} \left(
\begin{array}{cc}
 0 & \frac{M}{4 \hat{r}}-\frac{\hat{r}}{2} \\
 \frac{\hat{r}}{2}-\frac{M}{4 \hat{r}} & 0 \\
\end{array}
\right), \qquad e_x  = \left(
\begin{array}{cc}
 0 & \frac{\hat{r}}{2} \\
 \frac{\hat{r}}{2} & 0 \\
\end{array}
\right). \nonumber
\end{align}
Here,
\begin{align}
a_t & = \frac{\sqrt{M} \sinh 2 \sqrt{M} t}{2 \left(\cosh 2 \sqrt{M} t+\cosh 2 \sqrt{M} x\right)}\,,\quad
a_x  =\frac{\sqrt{M} \sinh 2 \sqrt{M} x}{2 \left(\cosh 2 \sqrt{M} t+\cosh 2 \sqrt{M} x\right)},\nonumber\\\nonumber\\
b & = \frac{1}{2} \sqrt{\frac{\cosh \sqrt{M}(t+x)}{\cosh \sqrt{M}(t-x)}}\,,\quad
c(x)  = \frac{M \left(3 \cosh \sqrt{M} (t-x)+\cosh \sqrt{M} (t+3 x)\right)}{4 \sqrt{2} \left(\cosh 2 \sqrt{M} t+\cosh 2 \sqrt{M} x\right)^{3/2}}, \nonumber\\
d(x)&=\frac{ \sqrt{M} \sinh \sqrt{M} (t+x)}{\sqrt{2}  \sqrt{ \cosh 2 \sqrt{M} t+\cosh 2 \sqrt{M} x}}.
\end{align}

\subsection{Conical defect to BTZ}
Let us consider the transformation from the conical defect metric of eq. \eqref{staticdeficit} to the BTZ metric. Using the transformation \eqref{AdS3toBTZ}, and the following connections in conical defect geometry,
\begin{align}
A^{\rm con}_r & =\frac{L_0}{\sqrt{r^2-T}}\,,\qquad
A^{\rm con}_\tau  \,= \,-\, \frac{L_{-1} T \exp (-\rho ) \,+\, L_{1} \exp (\rho )}{2 R}\,, \\ 
A^{con}_\phi & = \,-\, \frac{L_{-1} T \exp (-\rho ) \,+\, L_{1} \exp (\rho )}{2 R}\,, \nonumber 
\end{align}
where,
\begin{align}
T = \delta -R^2,
\end{align}
the following BTZ connection is obtained,
\begin{align}
A'_{\mu}(x') = \frac{\partial x^\nu }{\partial {x'}^\mu} A^{con}_\nu (x(x')).
\end{align}
Under the gauge transformation of equation \eqref{gauge_trans}, the above connection becomes the usual BTZ connection of \eqref{usualBTZconnec} plus a correction proportional to the deficit $\delta$,
\begin{align}
A^{\rm BTZ}\, =\, A^{(0)\rm BTZ} \,+\, \Delta A^{\rm BTZ},
\end{align}
where
\begin{align}\label{BTZ_A_corr}
\Delta A^{\rm BTZ} = \frac{-M \delta}{4 R^2 \hat{r}} (dt + dx) \big[ L_{1}  \text{sech}\left(\sqrt{M}(t+x)\right)\text{sech}\left(\sqrt{M}(t-x)\right)  \\
+ L_{-1} \text{sech}^2\left(\sqrt{M}(t+x)\right) \big] . \nonumber 
\end{align}

\subsection{Finite $\lambda$}
The following transformation between the set of coordinates, $x=(\tau,r,\phi)$ and $x'=(t,\hat{r},x)$, is used when $\lambda \neq 0$,
\begin{align}\label{coord_trans}
\sqrt{r^2+R^2}\,\sin \tilde{\tau} &\, =\, \frac{R}{\sqrt{M}}\,\sqrt{\hat{r}^2-M}\,\sinh\left(\sqrt{M}t\right)\,,\\\nonumber\\
\sqrt{r^2+R^2}\,\cos \tilde{\tau} & \,=\, \frac{R}{\sqrt{M}}{ \left[\hat{r}\cosh(\lambda)\cosh\left(\sqrt{M}x\right)-\sqrt{\hat{r}^2-M}\sinh(\lambda)\cosh\left(\sqrt{M}t\right) \right]}\,,\nonumber\\\nonumber\\
r \sin(\phi) & \,=\, \frac{R}{\sqrt{M}}\hat{r}\sinh\left(\sqrt{M}x\right)\,,\nonumber\\\nonumber\\
r\cos(\phi)& \,=\, \frac{R}{\sqrt M }{ \left[\cosh(\lambda)\sqrt{\hat{r}^2-M}\,\cosh\left(\sqrt{M}t\right)\,-\,\hat{r}\sinh(\lambda)\cosh\left(\sqrt{M}x\right) \right]}\,.\nonumber
\end{align}
The gauge transformation is of the form \eqref{gauge_trans_matrix}, where the components of the matrix are,
\begin{align}
u_{11}^{(0)} & =\sqrt{2b}, \qquad u_{22}^{(0)}=\frac{1}{u_{11}^{(0)}},\qquad
u_{12}^{(1)}  = u_{11}^{(0)} (a_t+a_x)\,,\\
u_{21}^{(1)} & =\frac{a_t-a_x}{ u_{11}^{(0)}}\,,\qquad
u_{11}^{(2)}  =\frac{2u_{11}^{(0)}(a_x u_{11}^{(0)} u_{12}^{(1)}+b u_{12}^{(1)} u_{21}^{(1)}-a_x d(x))}{2b+{u_{11}^{(0)}}^2}\,,\nonumber\\
u_{2}^{(2)} & =\frac{- 2 a_x u_{11}^{(0)} u_{12}^{(1)}+{u_{11}^{(0)}}^2 u_{12}^{(1)} u_{21}^{(1)}+ 2a_x d(x)}{2b u_{11}^{(0)}+ {u_{11}^{(0)}}^3}\,.\nonumber
\end{align}
where,
\begin{align}
a_t &= \frac{\sqrt{M} \cosh \lambda  \sinh \sqrt{M} t \left(\cosh \lambda  \cosh \sqrt{M} t-\sinh \lambda  \cosh \sqrt{M} x\right)}{2 \left(\left(\cosh \lambda  \cosh \sqrt{M} t-\sinh \lambda  \cosh\sqrt{M} x\right)^2+\sinh ^2\sqrt{M} x\right)},\\
a_x &= \frac{\sqrt{M} \cosh \lambda  \sinh \sqrt{M} x \left(\cosh \lambda  \cosh \sqrt{M} x-\sinh \lambda  \cosh \sqrt{M} t\right)}{2 \left(\left(\cosh \lambda  \cosh \sqrt{M} t-\sinh \lambda  \cosh \sqrt{M} x\right)^2+\sinh ^2\sqrt{M} x\right)}, \nonumber\\
b &= \tfrac{2^{-\frac{1}{2}}\cosh \lambda  \cosh \sqrt{M} (t+x)-\sinh \lambda }{ \left(\cosh 2 \lambda +\cosh ^2\lambda  \left(\cosh 2 \sqrt{M} t+\cosh 2 \sqrt{M} x\right)-2 \sinh 2 \lambda  \cosh \sqrt{M} t \cosh \sqrt{M} x-1\right)^{\frac{1}{2}}}, \nonumber\\
d(x) & = \tfrac{2^{-\frac{1}{2}}\sqrt{M}\cosh \lambda \sinh\sqrt{M} (t+x)}{\left(\cosh 2 \sqrt{M} t+\cosh 2 \sqrt{M} x\right) \cosh ^2\lambda +\cosh 2 \lambda -2 \cosh \sqrt{M} t \cosh \sqrt{M} x \sinh 2 \lambda -1}.\nonumber
\end{align}
The dreibeins $e'_t$, $e'_x$ and $e'_{\hat{r}}$ are as in equation \eqref{veir_BTZ_prime}.

\section{Conical defect with higher spin chemical potential}
In this apendix we study various aspects of a conical defect geometry with chemical potential for spin-3 charge.
\subsection{Free energy}
\begin{figure}[h]
\centering
\includegraphics[width=2.5in]{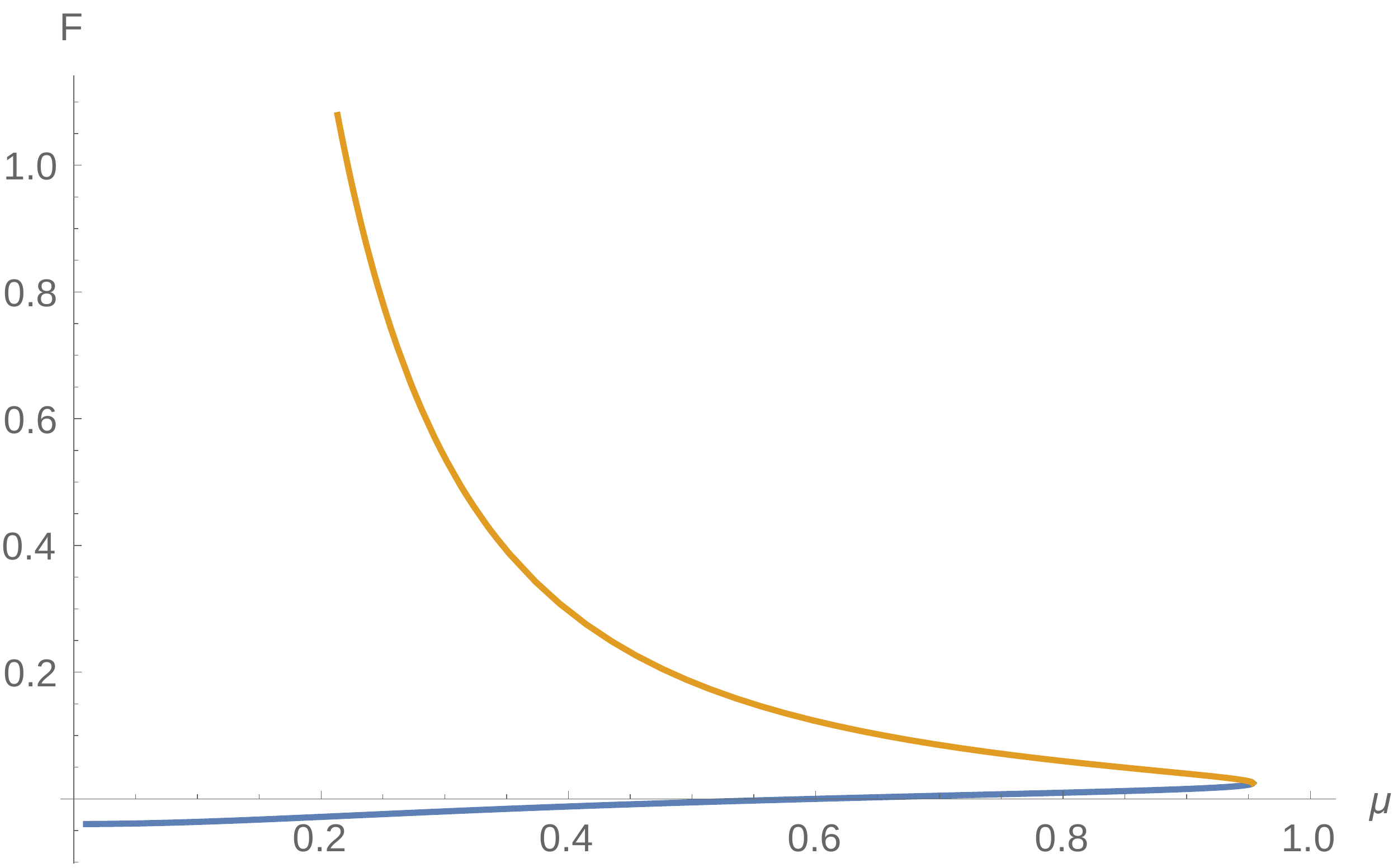}\hspace{0.5in}\includegraphics[width=2.5in]{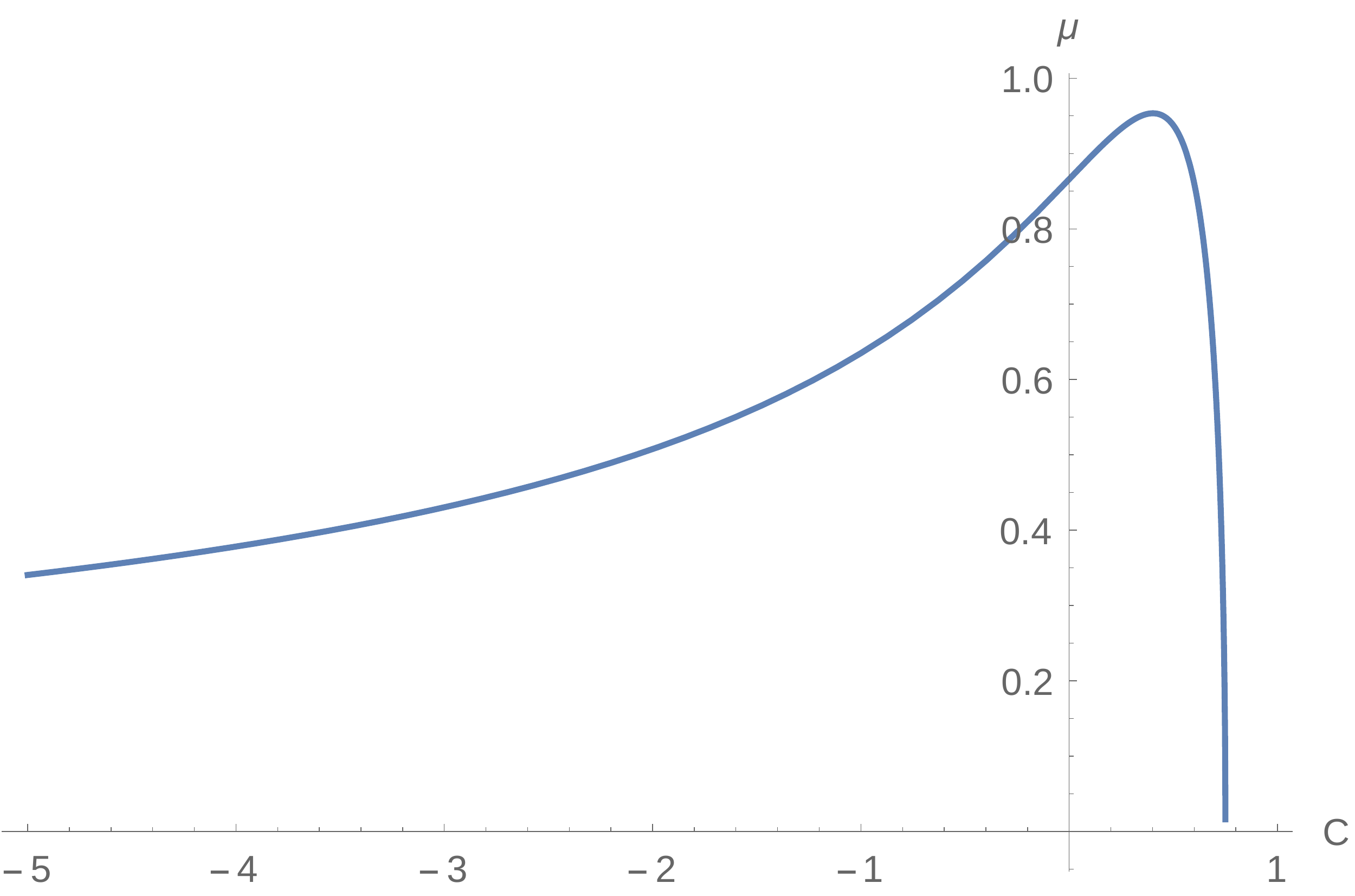}
\caption{\small {\bf Left:} Plot of free energy for the two sets of solutions of holonomy equations. The free energy is exact in $\mu$, and is plotted for $k=1$. {\bf Right:}Plot of $\mu$ versus $C$. $\mu$ is a single valued function for $C = (-\infty, \, \frac{3}{2} \left(2-\sqrt{3}\right)]$. For these values, $\mu = (0,\, \frac{-3}{8}  \sqrt{3+2 \sqrt{3}}]$.}
\label{free_en_plot}
\end{figure}

The following are the holonomy conditions for conical defect with higher spin chemical potential,
\begin{align} \label{phi_hol}
{\rm Tr} \left( a_\phi^2\right)\, =\, -2\left(1-\frac{\delta}{R^2}\right)\,, \qquad \det \left( a_\phi \right)\,=\,0\,,
\end{align}
where the eigenvalues of holonomy along the $\phi$ direction are,
\begin{align}
{\rm eval}(\omega_\phi)\, = \,\left(0,\,\pm 2\pi i \frac{\sqrt{R^2-\delta}}{R}\right)\,.
\end{align}
Solving these holonomy conditions gives two sets of solutions for ${{\cal L}, {\cal W}}$. The solution in equation \eqref{def_L_W_mu_CD} is chosen because it has lower free energy. The free energy is evaluated using 
\begin{align}
F = {\cal L} -\mu {\cal W}.
\end{align}
Figure \ref{free_en_plot} gives the plot of free energy for the two sets of solutions of holonomy equations. The blue curve with lower free energy corresponds to solution in equation \eqref{def_L_W_mu_CD}. In this plot, the free energy is exact in $\mu$, and is plotted for $k=1$.

\subsection{Entanglement entropy for static case}
The entanglement entropy for conical defect with higher spin chemical potential is, 
\begin{align}
S_A& = \frac{c}{3} \log \left(\frac{2\Lambda \sin\left(\frac{\sqrt{R^2-\delta}}{R}\frac{\phi}{2}\right)}{\sqrt{R^2-\delta}}\right)\\
& + \frac{c}{72} \frac{\mu^2}{R^2} \frac{R^2-\delta}{R^2} \text{cosec}^4\left(\frac{\sqrt{R^2-\delta}}{R}\frac{\phi}{2}\right) \left[3-2\left(3 \frac{R^2-\delta}{R^2} \phi ^2+4\right) \cos \left(\frac{\sqrt{R^2-\delta}}{R} \phi \right) \right.\nonumber\\
& \left. +4 \frac{\sqrt{R^2-\delta}}{R} \phi  \left(\sin \left(\frac{\sqrt{R^2-\delta}}{R} \phi \right)+\sin \left(2\frac{\sqrt{R^2-\delta}}{R} \phi \right)\right)+5 \cos \left(2 \frac{\sqrt{R^2-\delta}}{R} \phi \right)\right]\nonumber\,.
\end{align}
Here, $\tau_1 = \tau_2 =0$ and $\phi = (\phi_2-\phi_1)/2$.
The $\delta \to 0$ limit of this result matches with entanglement entropy of AdS$_3$ with higher spin chemical potential given in \cite{Datta:2014ypa}. Here $\Lambda$ is the UV cutoff.

\subsection{Solving holonomy equations in terms of $C$}
Solving the $\phi$ holonomy conditions of equation \eqref{phi_hol}, the following equations are obtained, where for convenience we use $k=4k_{\rm cs}$,
\bea
&& 256 \pi^2 \mu^2 {\cal L}^2 +24 \pi k {\cal L} + 72 \pi k \mu {\cal W} +3k^2 =0\,,\\\nonumber
&&2048 \pi^2 \mu^3 {\cal L}^3 - 576 \pi k \mu {\cal L}^2 - 864 \pi k \mu^2 {\cal W} {\cal L} -864 \pi k \mu^3 {\cal W}^2 -27 k^2 {\cal W}=0\,.
\eea
We define the parameters,
\bea
&& \zeta =\sqrt{\frac{k}{32 \pi {\cal L}^3}}{\cal W}, \qquad \gamma= \sqrt{\frac{2\pi{\cal L}}{k}}\mu\,,
\eea
to write the holonomy conditions as,
\bea\label{phi_hol_reparam}
&& 1728 \gamma^3 \zeta^2+(432\gamma^2+27)\zeta-128\gamma^3+72\gamma=0,\\
&&\frac{k}{8\pi}+(1+\frac{16}{3}\gamma^2+12\gamma\zeta){\cal L}=0. \nonumber
\eea
Substituting,
\begin{align}
\zeta = \frac{1-C}{C^{3/2}},
\end{align}
in the first of the above equations, the expression for $\gamma$ is,
\begin{align}
\gamma = \frac{\sqrt{C}}{4(2C-3)}.
\end{align}
Thus,
\begin{align}
{\cal W} = \frac{4(1-C)}{C^{3/2}}{\cal L} \sqrt{\frac{2\pi{\cal L}}{k}}, \quad \mu = \frac{\sqrt{C}}{4(2C-3)} \sqrt{\frac{k}{2\pi{\cal L}}},
\end{align}
and using the second equation of equation \eqref{phi_hol_reparam}, $\mu$ is obtained entirely as a function of $C$,
\begin{align}\label{mu_def_c}
\mu =- \frac{3(C-3)\sqrt{3-4C}}{2(3-2C)^2}.
\end{align}
$\mu$ is a single valued function for $C = (-\infty, \, \frac{3}{2} \left(2-\sqrt{3}\right)]$. For these values, $\mu$ runs from $\mu = (0,\, \frac{-3}{8}  \sqrt{3+2 \sqrt{3}}]$, that is roughly, $\mu \sim (0, \, -0.953]$. Figure \ref{free_en_plot} gives the plot for $\mu$ versus $C$. 
%Using the following,
%\begin{align}
%\xi_1 &=-\frac{\sqrt[3]{\frac{2}{3}} \left(\left(\frac{3}{2}\right)^{2/3} \left(3 \left(\sqrt{9-12 C}+3\right)-\left(\sqrt{9-12 C}+9\right) C\right)^{2/3}+\frac{3 C (3-2 C)}{C-3}\right)}{3 \sqrt{3-4 C} \sqrt[3]{3 \left(\sqrt{9-12 C}+3\right)-\left(\sqrt{9-12 C}+9\right) C}}\\
%\xi_2 &=-i\frac{\sqrt[3]{\frac{2}{3}} \left(-\left(\frac{3}{2}\right)^{2/3} \left(3 \left(\sqrt{9-12 C}+3\right)-\left(\sqrt{9-12 C}+9\right) C\right)^{2/3}+\frac{3 C (3-2 C)}{C-3}\right)}{3 \sqrt{3-4 C} \sqrt[3]{3 \left(\sqrt{9-12 C}+3\right)-\left(\sqrt{9-12 C}+9\right) C}}\nonumber
%\end{align} 
%in equaation \eqref{WL_condefect} and using,
%\begin{align}
%S_A = \frac{c}{48} \log \left( W_{Adj} \right),
%\end{align}
%entanglement entropy as a function of $C$ is obtained. Using equation \eqref{mu_def_c}, the entanglement entropy as a function of $\mu$ is obtained. 

%\begin{figure} 
%\centering
%\includegraphics[scale=0.5]{mu_vs_c_plot.pdf}
%\caption{Plot of $\mu$ versus $C$. $\mu$ is a single valued function for $C = (-\infty, \, \frac{3}{2} \left(2-\sqrt{3}\right)]$. For these values, $\mu = (0,\, \frac{-3}{8}  \sqrt{3+2 \sqrt{3}}]$.}
%\label{mu_vs_c}
%\end{figure}
The relation between $\mu$ and $C$ for conical defect with higher spin chemical potential is given by,
\begin{align}\label{mu_def_c}
\frac{\mu}{R} =- \frac{3(C-3)\sqrt{3-4C}}{2\sqrt{1-\delta}(3-2C)^2}\,.
\end{align}

\section{Another setup for infalling particle in eternal BTZ}
\label{app:infallingCS}
This section follows the analysis of \cite{Shenker:2013pqa}. Consider the Kruskal extension of BTZ blackhole, the BTZ metric transforms to the following metric in light cone coordinates,
\begin{align}
ds^2=\frac{-4R^2 du dv + r_H^2(1-uv)^2d\phi^2}{(1+uv)^2},
\end{align}
using the transformation,
\begin{align} \label{Kruskal_trans_sl2}
t= \frac{\beta}{4\pi}\log\left(\frac{-v}{u}\right), \qquad r=r_H\frac{1-uv}{1+uv}.
\end{align}
The right region is $(u<0,v>0)$ and the left region is $(u>0,v<0)$. The left asymptotic region can be reached from right asymptotic region by translating the time as $t \to t+ i\beta/2$.

We want to study this geometry when a massive particle is added at time $-t_w$ at the left boundary and it falls towards the horizon. This gives rise to a backreacted geometry. This backreacted geometry is modelled by gluing together a BTZ solution of mass $M$ (right region) to a BTZ solution of mass $M+E$ (left region), along the null surface $u_w=e^{-\frac{2\pi t_w}{R^2}}$. Here $E$ is the energy of the infalling perturbation and is considered to be much smaller than $M$. The radius of horizon is different in the two regions and is related by,
\begin{align}
\tilde{r}_H = \sqrt{\frac{M+E}{M}}r_H.
\end{align}
The right region has the coordinates $(u,v)$, and left region has coordinates $(\tilde{u},\tilde{v})$. The following matching conditions relate these sets of coordinates:
\begin{itemize}
\item $\tilde u_w=u_w=e^{-\frac{2\pi t_w}{l^2}}$. This is true in the limit $E/M \to 0$ followed by $t_w \to \infty$.
\item The coefficient of $d\phi^2$ in the metric is continuous across the shell. 
%$\langle$ However we want to make the following metric independent statement: The closed Wilson line along the $\phi$ direction is continuous across the shell. $\rangle$
\begin{align}
\tilde{r}_H \frac{1-\tilde{u}\tilde{v}}{1+\tilde{u}\tilde{v}} = r_H \frac{1-u v}{1+u v}.
\end{align}
\end{itemize}
Solving this equation in the limit $E/M \to 0$,
\begin{align}
\tilde{v} = v + \frac{E}{4M}\left(e^{\frac{2\pi t_w}{\beta}}-v^2e^{\frac{-2\pi t_w}{\beta}}\right),
\end{align}
next the limit $t_w \to \infty$ is taken,
\begin{align}\label{def_alpha}
\tilde{v} = v+\alpha, \quad \alpha = \frac{E}{4M}e^{\frac{2\pi t_w}{\beta}}.
\end{align}
We now proceed to evaluate the Wilson line from a point $(\rho_P, t_L, \phi)$ at the left boundary to a point $(\rho_Q, t_R, \phi)$ at the right boundary. In order to do this we first evaluate the geodesic length from the point $(\rho_Q,t_R, \phi)$ at the right boundary to an intermediate point $(\rho_S,t_S, \phi)$ at the shell at $u_w$. This is obtained by taking, $\frac{2\pi {\cal L}}{k_{cs}} = \frac{2\pi \bar{{\cal L}}}{k_{cs}} = \frac{\pi^2 R^2}{\beta^2}$, and $\rho_Q \to \infty$ in the Wilson line of equation \eqref{sl2wl},
\begin{align}
W^R_{\rm fund} = \frac{r_\infty}{R} \left[ e^{-\rho_S} \cosh \left( \frac{\pi}{\beta}\Delta \xi^+ \right) \cosh \left( \frac{\pi}{\beta} \Delta \xi^- \right) - \frac{e^{\rho_S} \beta^2}{\pi^2 R^2} \sinh \left(  \frac{\pi}{\beta}\Delta \xi^+ \right)\sin \left(  \frac{\pi}{\beta}\Delta \xi^- \right) \right],
\end{align}
%\begin{align}
%{\cal W}_{SL(2)}=\frac{e^{-\frac{\pi (t_1+t_2)}{\beta }-\rho_2} \left(\pi ^2 l^2 \left(e^{\frac{\pi t_1}{\beta }}+e^{\frac{\pi t_2}{\beta }}\right)^2-\beta ^2 e^{2 \rho_2} \left(e^{\frac{\pi t_1}{\beta }}-e^{\frac{\pi t_2}{\beta }}\right)^2\right)}{4 \pi ^2 l^2}.
%\end{align}
where, $\xi^\pm = t \pm R \phi$. Using the transformation of equation \eqref{Kruskal_trans_sl2} and the following equation,
\begin{align}
\rho =  \log \left[ \frac{1}{2R}\left(r+\sqrt{r^2-r^2_H}\right)\right],
\end{align}
$(\rho_S,t_S)$ is substituted in terms of $(u_w,v)$. Taking the limit $u_w \to 0$, we obtain,
\begin{align} \label{WL_sl2_R}
\log W^R_{\rm fund} = \log\left( \frac{r_\infty \beta}{\pi R^2} \right)+\log\left(1-v \, e^{-\frac{2\pi t_R}{\beta}} \right).
\end{align}
The Wilson line from the point at the left boundary $(\rho_P,t_L, \phi)$ to the point at the shell $(\rho_S,t_S, \phi)$, is obtained by performing the following shift in equation \eqref{WL_sl2_R}: $v \to v+\alpha $, and $t_R \to t_L+\frac{i\beta}{2}$,
\begin{align} \label{WL_sl2_L}
\log W^L_{\rm fund} = \log\frac{r_\infty \beta}{\pi R^2}+\log\left(1+(v+\alpha) e^{-\frac{2\pi t_L}{\beta}} \right).
\end{align}
To obtain the Wilson line from the left to right boundary, minimize the function $\log \left( W^R_{\rm fund} W^L_{\rm fund}\right)$ over
$v$. The value of $v$ at the minimum is,
\begin{align}\label{v_min}
v_{min} = \frac{1}{2}\left(-e^{\frac{2\pi t_L}{\beta}}+e^{\frac{2\pi t_R}{\beta}} -\alpha\right) .
\end{align}
Thus, the Wilson line from the left to right boundary is,
\begin{align}
\log  W_{\rm fund} & = \log \left( W^R_{\rm fund} W^L_{\rm fund}\right)|_{v_{min}}\\
& = 2\log\frac{r_\infty \beta}{\pi R^2} + 2 \log\left(\frac{1}{2}e^{-\frac{4\pi}{\beta}(t_L+t_R)} \left(e^{\frac{2\pi}{\beta}t_L}+e^{\frac{2\pi}{\beta}t_R} +\alpha\right) \right).\nonumber
\end{align}
The mutual information is evaluated at $t_L=t_R=0$, and the scrambling time $t^*_w$ is obtained by setting mutual information to zero, $I(A;B)(t^*_w)=0.$ 
The mutual information is,
\begin{align}
I(A;B) =\log \sinh \frac{\pi R \phi}{\beta} -\log\left(1+\frac{E\beta}{4S}e^{\frac{2\pi t_w}{\beta}} \right), \nonumber
\end{align}
where, $S_A,S_B$ are entanglement entropies of regions of size $R\phi$ on the two boundaries,
\begin{align}
S_A = S_B = \frac{1}{4} \left( 2\log\frac{r_\infty \beta}{\pi R^2}+2\log \sinh \frac{\pi R \phi}{\beta} \right),\\
S_{A \cup B} = \frac{1}{2} \log W_{\rm fund} = \log\frac{r_\infty \beta}{\pi R^2}+\log\left(1+\frac{\alpha}{2}\right).\nonumber
\end{align}
Thus the scrambling time is,
\begin{align}
t^*_w = \frac{\phi R}{2}+\frac{\beta}{2\pi} \log \left( \frac{2S}{\beta E} \right).
\end{align}
%\end{appendix}
%\bibliography{scrambling}
%\bibliographystyle{JHEP}

\end{document}